\documentclass[twocolumn, prl, superscriptaddress]{revtex4-2}

\usepackage{color,soul}
\usepackage{times}
\usepackage{graphicx}
\usepackage{soul}
\usepackage{amsmath}
\usepackage{amssymb}
\usepackage{units}
\usepackage{ulem}
\usepackage{cancel}
\usepackage{babel}
\usepackage{hyperref}
\usepackage{units}
\newcommand{\abvec}{\left(\begin{array}{c} \hat{a}\\ \hat{b} \end{array}\right)}
\newcommand{\pt}{\partial_t}
\newcommand{\ha}{\hat{a}}

\newcommand{\hb}{\hat{b}}

\newcommand{\ea}{\langle\hat{a}\rangle}

\newcommand{\eb}{\langle\hat{b}\rangle}

\newcommand{\ra}{\mathrm{a}}
\newcommand{\rb}{\mathrm{b}}
\newcommand{\rc}{\mathrm{c}}
\newcommand{\rp}{\mathrm{p}}

\newcommand{\tz}{\tilde{\zeta}}
\newcommand{\wZ}{\omega_\rb}
\newcommand{\ex}{\mathbf{e}_x}
\newcommand{\ey}{\mathbf{e}_y}
\newcommand{\ez}{\mathbf{e}_z}
\newcommand{\Methods}{Methods}
\newcommand{\HzUnit}{Hz}

\begin{document}

\title{Strong coupling of alkali-metal spins to noble-gas spins with an hour-long coherence time}

\author{R. Shaham}
\thanks{These authors contributed equally to this work.}
 \affiliation{Department of Physics of Complex Systems, Weizmann Institute of Science, Rehovot 76100, Israel}
 \affiliation{Rafael Ltd, IL-31021 Haifa, Israel}
\author{O. Katz}
\thanks{These authors contributed equally to this work.}
 \affiliation{Department of Physics of Complex Systems, Weizmann Institute of Science, Rehovot 76100, Israel}
 \affiliation{Rafael Ltd, IL-31021 Haifa, Israel}
 \affiliation{Present address: present address: Department of Electrical and Computer Engineering, Duke University, Durham, NC 27708}
\author{O. Firstenberg}
\affiliation{Department of Physics of Complex Systems, Weizmann Institute of Science, Rehovot 76100, Israel}

\begin{abstract}
Nuclear spins of noble gases can maintain coherence for hours at ambient conditions because they are isolated by complete electron shells \cite{Walker2017He3review}.
This isolation, however, impedes the ability to manipulate and control them by optical means or by coupling to other spin gases \cite{WalkerLarsen2016NGCNMRG,KornackRomalis2002OverlappingEnsemles,JimenezMartinez2014KitchingXeNMR}.
Here we achieve strong coherent coupling between noble-gas spins and the optically-accessible spins of an alkali-metal vapor.
The coupling emerges from the coherent accumulation of stochastic spin-exchange collisions.
We obtain a coupling strength ten times higher than the decay rate, observe the coherent and periodic exchange of spin excitations between the two gases, and demonstrate active control over the coupling by an external magnetic field.
This approach could be developed into a fast and efficient interface for noble-gas spins, enabling applications in quantum sensing and information \cite{AlkaliNobleEntanglementKatz2020PRL,noblegasStoragePRL2020arxiv}.
\end{abstract}

\maketitle

\section* {Introduction}
Noble gas isotopes with a nonzero nuclear spin, such as helium-3, feature day-long spin lifetimes and hours-long coherence times \cite{Heil2013T2100h,Walker2017He3review}.
They are prominent in various fields, from precision sensing \cite{gemmel2010UltraSensitiveMagnetometer,Kornack2005GyroComagRomalis,Thrasher2019DualSpeciesSynchronousSEOP,Kitching2018atomicDevices} and medical imaging \cite{Chupp2001Imaging} to searches of new physics \cite{Brown2010RomalisCPTviolation,JacksonKimball2010BudkerConstraintsNaHe,Alonso2019DarkMatterComagClocks,Bloch2019Axions,Chupp2019EDMreview}, and they hold promise for future quantum information applications such as optical quantum memories and the generation of long-lived entanglement \cite{AlkaliNobleEntanglementKatz2020PRL,noblegasStoragePRA2020arxiv,Dantan2005SinatraPinardMEOPstorage,SerafinSinatra2020MEOPsqueezing,SerafinSinatra2021QuantumMEOPfull}.
The latter rely on the feasibility of preparing the collective spin state of the gas and controlling its quantum excitations \cite{weakcollisions2019arxiv}.

Polarized ensembles of alkali-metal spins or noble-gas spins can carry such collective excitations, corresponding classically to a tilt of the collective spin about the polarization axis \cite{Polzik2010ReviewRMP}.
These can be modeled as quantum excitations of a harmonic oscillator. 
Remarkably, the quantum description persists even for gaseous ensembles undergoing rapid diffusion \cite{Shaham2020Diffusion,Xiao2019MultiplexingSqueezedLightDiffusion} and for overlapping ensembles that interact via atomic collisions \cite{weakcollisions2019arxiv,Dellis2014SESpinNoisePRA,Kong2018MitchellAlkaliSEEntanglement,Mouloudakis2020SEbipartiteEntanglement}.
The collective state of alkali spins can be addressed and coherently controlled by optical means \cite{Sherson2006PolzikTeleportationDemo,Gorshkov2007UniversalOptimalStoragePRL,firstenberg2010self}.
The same, however, cannot be done for the nuclear spins of noble gases, which lack any optical transition from the ground levels.
Instead, one can access the noble-gas spins by collisions with another spin gas, either excited (metastable) helium-3 or alkali-metal vapor, both of which possess optically-accessible spins
\cite{AppeltHapper1998SEOPtheoryPRA,Walker2017He3review,Batz2011MEOPreview}. 

Alkali metal atoms exchange spin with noble-gas atoms via a weak electron-nuclear coupling (Fermi contact) during collisions \cite{Walter1998HapperWalkerPhiTrajectory}.
They are normally used for hyperpolarizing the noble gas and for probing its spin dynamics.
The probing relies on the coherent component of the spin-exchange interaction, which is usually weak and manifests as a shift in the precession frequencies of the alkali spins.
The coherent component is employed for the readout of noble-gas-based sensors and for inherent suppression of sensitivity to magnetic fields \cite{WalkerLarsen2016NGCNMRG,KornackRomalis2002OverlappingEnsemles,Kornack2005GyroComagRomalis,JimenezMartinez2014KitchingXeNMR}.

In Ref.~\cite{KornackRomalis2002OverlappingEnsemles}, Kornack and Romalis employ the coherent component of the spin-exchange interaction to study the hybridization of the collective spins of alkali-metal and noble-gas ensembles at the critical-damping regime.
By varying the axial magnetic field, they observe shortening of the noble-gas spin coherence time and shifting of the noble-gas magnetic resonance due to the alkali-spin dressing.
With a coupling rate lower than the alkali decay rate, they observe the onset of avoided crossing in the magnetic spectrum but do not record reversible dynamics or revivals of the spin.
In \cite{noblegasSpectroscopy2020arxiv} we report on using alkali spins as off-resonant mediators to couple light to noble-gas spins bidirectionally.
These and all other works so far have been limited to the detuned or critical coupling regimes.

Increasing the coupling rate between the spin gases is beneficial to various applications.
When the coupling exceeds the alkali decay rate, the dynamics of the spin gases become strongly coupled, enabling rapid and coherent control of the noble-gas spins.
Operation in this strong-coupling regime opens new practical avenues, especially for in-and-out mapping of quantum states, for enhancing the indirect interaction of noble-gas spins with photons, and for improving the performance of sensing applications \cite{weakcollisions2019arxiv,noblegasStoragePRA2020arxiv,noblegasStoragePRL2020arxiv}.
Yet, strongly coupled dynamics and, in particular, the reversible exchange between the spin gases have never been demonstrated.

Here we report on the strong coherent coupling between the collective spin states of noble-gas and alkali-metal ensembles.
We enter the strong-coupling regime by reaching high polarizations and densities of the interacting species while minimizing spin relaxation.
We directly probe the dynamics of both spin ensembles and demonstrate the coherent and reversible exchange of excitations between them.
These results demonstrate that stochastic spin-exchange collisions which are individually weak but altogether frequent enough can accumulate to form an efficient coherent interface between two spin gases.
We discuss prospects for strongly-coupled gases and their potential utility in classical and quantum applications.

To characterize the exchange, we consider the bosonic collective spin excitations of the alkali-metal and noble-gas spins, represented by the annihilation operators $\hat{a}$ and $\hat{b}$, respectively.
The coupling between these excitations relies on the collective enhancement of the exchange interaction, due to accumulation of numerous collisions among the two spin ensembles.
The collective, bi-directional, coupling rate $J=(\tz/2) \sqrt{n_\ra p_\ra n_\rb p_\rb}$ thus depends on the square root of the atomic densities $n_\ra$, $n_\rb$ and degrees of polarization $0\le p_\ra,p_\rb\le 1$ \cite{weakcollisions2019arxiv}.
The microscopic coupling strength $\tz (p_\ra)$, incorporating the collisional cross-section, has a weak dependence on the alkali spin polarization due to the hyperfine structure of the alkali atoms (see \Methods{}).
A simple form of two coupled modes can be used to describe the exchange dynamics,
\begin{equation} 
\pt \abvec = i \left( \begin{array}{cc}
\omega_\ra+i\gamma & -J \\ -J & \wZ
\end{array} \right) \abvec + \hat{\boldsymbol{f}}.
\label{eq:coupled-spins} 
\end{equation}
Here $\omega_\ra$ and $\wZ$ denote the Larmor precession frequencies of the collective spins of the alkali and noble-gas atoms, respectively.
They are set by the external magnetic field $B$ and by the effective magnetic fields exerted by each species on the other \cite{AppeltHapper1998SEOPtheoryPRA}.
We tune $B$ to determine the detuning from resonant coupling $\Delta=\omega_\ra-\wZ$.
The decoherence rate of the alkali excitations $\gamma$ is included, while for now we neglect the slow decoherence of the noble-gas spins.
Finally, $\hat{\boldsymbol{f}}$ denotes the quantum noise accompanying the relaxation, motion, and collision processes \cite{weakcollisions2019arxiv,Shaham2020Diffusion}.
In the current study, $\hat{\boldsymbol{f}}$ can be discarded, as we prepare the spin ensembles in coherent spin states and study the evolution of the mean transverse amplitudes $\ea$ and $\eb$.

\section* {Results}
\subsection*{Experimental setup and protocols}
\begin{figure}[tb]
\centering
\includegraphics[width=0.95\columnwidth]{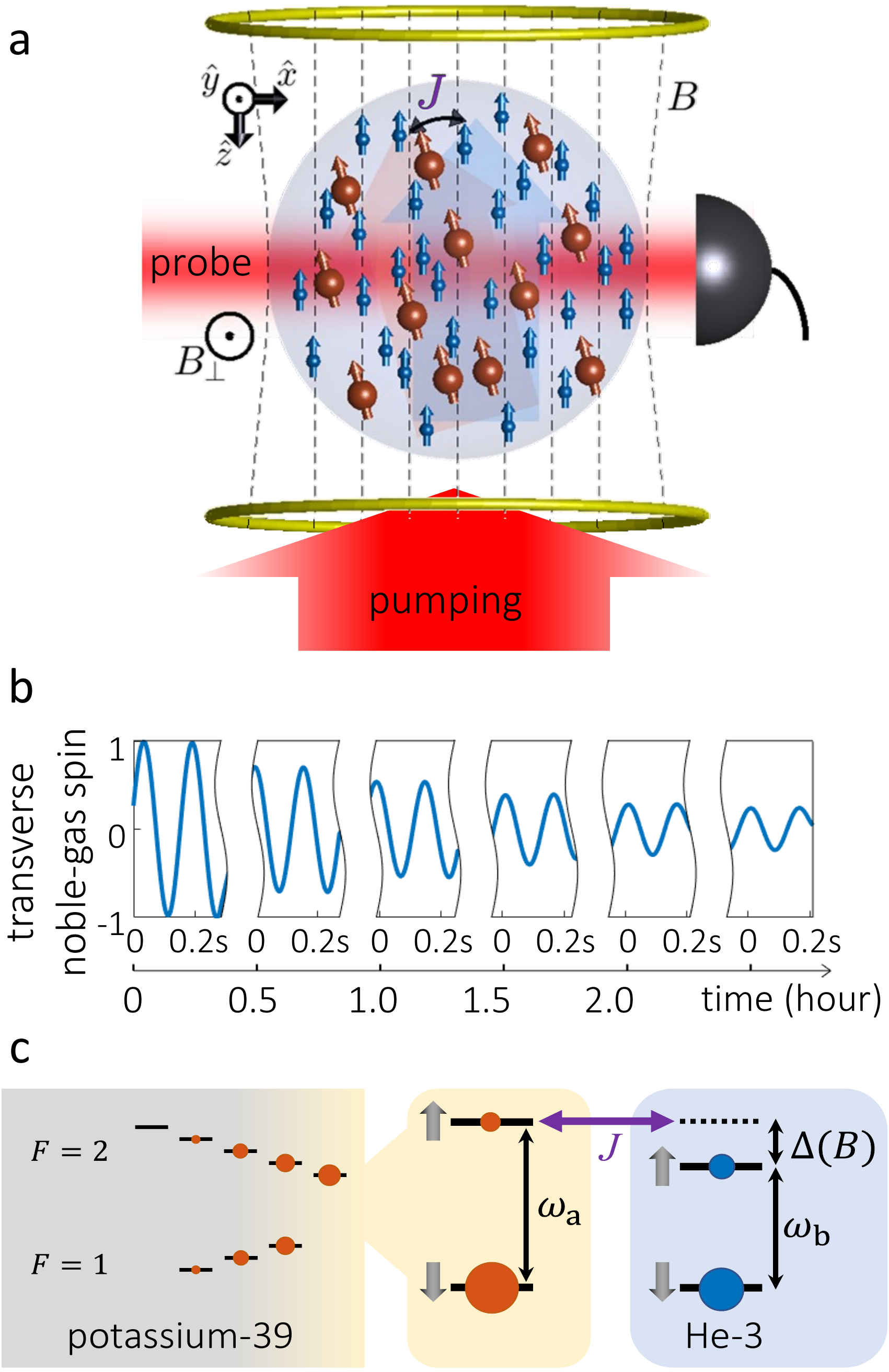}
\caption{\label{fig:apparatus}
\textbf{Experimental scheme and coherence-time measurements.}
\textbf{a.} A glass cell containing optically-pumped potassium vapor (alkali-metal spins, red) and helium-3 (noble-gas spins, blue).
The polarized ensembles couple via stochastic atomic  collisions that accumulate to a collective spin-exchange interaction at a rate $J$.
An applied magnetic field $B\hat{z}$ controls the precession frequency difference $\Delta=\omega_\ra-\omega_\rb$ between the two ensembles.
A transverse excitation of the spins is initialized by a short transverse magnetic field pulse $B_\perp\hat{y}$ and then monitored by Faraday rotation of an optical probe.
\textbf{b.} Precession of the helium-3 spins, measured at low spin polarizations and normalized to the initial value, featuring a coherence time of $T^\rb_2=2$ hours.
\textbf{c.} Energy level diagram for the coupled spins.
The spin-polarized alkali atoms, undergoing frequent spin-exchange collisions, can be described as an effective two-level system.
}
\end{figure}

We study transverse spin excitations of polarized potassium vapor and helium-3 gas enclosed in a spherical glass cell, as shown in Fig.~\ref{fig:apparatus}a.
The potassium spins are polarized along the axial magnetic field by an optical pumping beam, and the helium spins are polarized by collisions with the spin-polarized potassium (over 10 hours, see Fig.~\ref{fig:SEOP}). The cell also contains nitrogen for reducing (quenching) the fluorescence from the optically excited potassium atoms.
At low polarizations, the helium spins exhibit a coherence time of $T_2^{\rb}=2$ hours, as shown in Fig.~\ref{fig:apparatus}b, and consequently their individual relaxation is henceforth neglected.
The exchange experiments start straight after turning off the pumping beam.

We monitor the dynamics of the coupled spin system following a 5-$\mu$s-long pulse of transverse magnetic field $B_\perp$, which predominantly excites the collective alkali spin and initializes it at a tilt angle of a few degrees from the axial magnetic field $B\hat{z}$.
We measure the transverse alkali spin using Faraday rotation of an optically-detuned, linearly-polarized probe beam.
In this system, the exchange rate $J$ and the magnetic precession rates $\omega_\ra$ and $\omega_\rb$ are all of the same scale when $\Delta\lesssim J$.
As a result, strongly-coupled dynamics measured in the lab frame mix the effects of Larmor precession with that of the exchange.
To eliminate the effect of the former and witness the exchange dynamics directly, we experimentally reconstruct the complex quantities $\ea$ and $\eb$, each composed of the two spin components (quadratures) in the transverse $xy$ plane.
The tomographic-like reconstruction is performed by repeated measurements of the alkali spin dynamics in the $xy$ plane using alternated pulses $B_\perp\hat{y}$ and $B_\perp\hat{x}$ for the initial tilt.
We properly scale these measurements by the total degree of polarization $p_\ra(t)$ (measured independently, see Fig.~\ref{fig:PandQ}) and calculate the complex amplitude of the collective alkali spin $\langle \hat{a}(t)\rangle$.
To measure the collective noble-gas spin $\langle\hat{b}(t)\rangle$ after some exchange duration $t$, we halt the exchange dynamics at $t$ by rapidly ramping up the axial magnetic field (increasing $\Delta$) and utilizing the alkali spins as a magnetometer for sensing the noble-gas spin precession.

We realize a maximal coupling rate of $J=78\pm 8$ \HzUnit{}  by operating at high densities of potassium $n_\ra=4.9\cdot10^{14}/\mathrm{cm}^3$ (at $T=230~^\circ\mathrm{C}$) and helium $n_\rb=6.45\cdot10^{19}/\mathrm{cm}^3$ (2.4 atm at room temperature) and with relatively high degrees of spin polarization $p_\ra\gtrsim0.95$ and $p_\rb\gtrsim0.3$.
At these conditions, collisions among alkali atoms are frequent enough ($>0.5/\mathrm{\mu s}$) with respect to the Larmor frequency to keep the alkali excitations free from spin-exchange relaxation (so-called SERF regime) \cite{AppeltHapper1998SEOPtheoryPRA}.
The intricate hyperfine manifold of the alkali atoms maintains a spin-temperature distribution due to these collisions and manifests as an effective spin-1/2, as shown in Fig.~\ref{fig:apparatus}c \cite{AppeltHapper1998SEOPtheoryPRA}.
Remnant spin-relaxation occurring during atomic collisions dominate the decoherence rate of the alkali excitations $\gamma = 7.3 \pm 1.5$ \HzUnit{}.
We thus achieve $J \gtrsim 10 \gamma$.
See \Methods{} for a detailed description of the experimental conditions and analysis procedures.

\subsection*{Dynamics of strongly-coupled spins}

\begin{figure}[tb]
\centering
\includegraphics[width=1.0\columnwidth]{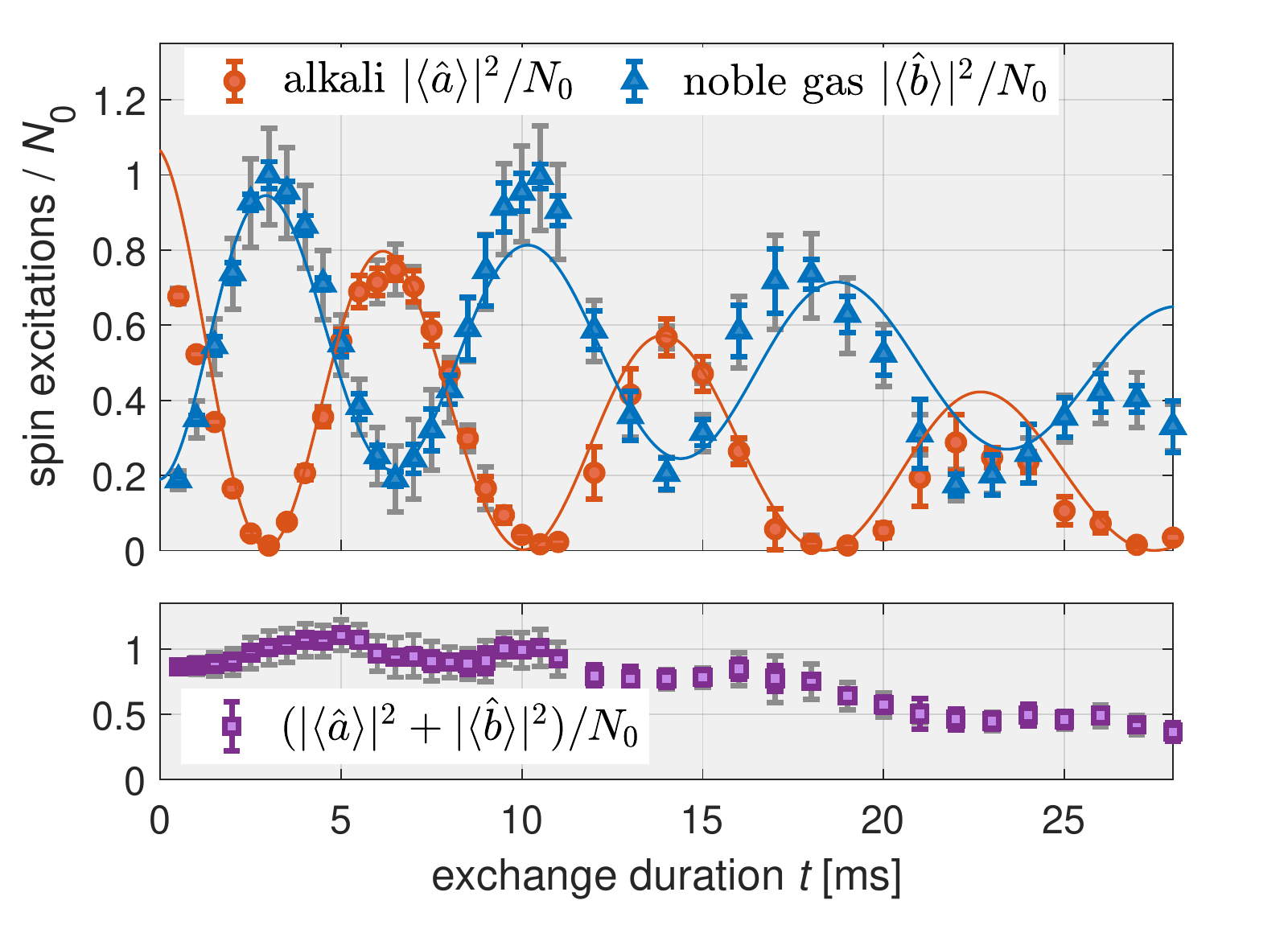}
\caption{
\textbf{Exchange of collective spin excitations.}
Measurement of the coherent exchange between the alkali spin $\langle\hat{a}\rangle$ (red circles) and the noble-gas spin $\langle\hat{b}\rangle$ (blue triangles) in the strong-coupling regime.
A short pulse of transverse magnetic field at $t=0$ excites $N_0 = |\langle\hat{a}(0)\rangle|^2 + |\langle\hat{b}(0)\rangle|^2 = (13.2\pm0.6) \cdot 10^{13}$ spins.
The experimental conditions at $t=0$ are $J=78\pm8$ \HzUnit{}, $\gamma=7.3\pm1.5$ \HzUnit{}, and $\Delta=-1.15J$ (for obtaining maximal extinction of $\langle\hat{a}\rangle$ at the minima, see text).
Lines present the result of a detailed model using these parameters, obtained from independent measurements.
Each data-point is averaged over 12 to 20 repetitions of the experimental sequence (shown in Fig.~\ref{fig:sequence}).
Colored errorbars include uncertainties in the spin-projection measurements and the scattering between repetitions.
Gray errorbars indicate uncertainty due to the uncertainty in the alkali polarization $p_\ra(t)$, required for converting spin projections to excitations.
The bottom panel presents the same data in terms of $|\langle\hat{a}\rangle|^2 + |\langle\hat{b}\rangle|^2$, confirming that the total number of excitations is conserved by the exchange process, up to an overall decoherence.
}
\label{fig:exchange}
\end{figure}

Under the strong-coupling conditions, the two spin gases can coherently exchange collective excitations.
To demonstrate these dynamics, we tune $\Delta$ close to resonance and generate an initial excitation predominantly of the alkali spin.
Figure~\ref{fig:exchange} presents the measured spin excitations  $|\langle\hat{a}\rangle|^2$ and $|\langle\hat{b}\rangle|^2$, as they are exchanged back and fourth between the two ensembles.
Because the magnetic pulse acts also on the noble-gas spin and partially excites it as well, the extinction of $|\langle\hat{a}(t)\rangle|^2$ at the minima of the observed oscillations is maximized slightly below resonance, at $\Delta=-1.15J$; the presented measurement is taken at this detuning.
This detuning is still small in terms of the strong-coupling dynamics, rendering a near-unity ratio between the exchange and coupling rates $\tilde{J}/J=1.15$, where  $\tilde{J}\approx\sqrt{J^2+\Delta^2/4}$ is the exchange rate.
We observe the reversible exchange and the revival of the excitations back to the alkali spins at $t=6.5$ ms and with a high contrast of 75\%, evidencing the strongly coupled dynamics.
The uncertainty on the exchange fidelity is small ($7.4\%$) for short exchange times, and it is dominated by uncertainty on the alkali polarization (see Methods).

We find as expected that the exchange conserves the total number of excitations $|\langle\hat{a}\rangle|^2+|\langle\hat{b}\rangle|^2$ aside from the decay introduced by alkali-spin decoherence.
We also directly observe the slowing down of the exchange oscillations, as the spins gradually decouple due to the dependence of $\tilde{J}$ on the decaying alkali polarization $p_\ra(t)$.
The gradual decoupling leads to residual excitations populating the long-lived noble gas spin.
These effects are all captured by a detailed model (solid lines),
described in \Methods{}, which accounts for the temporal decrease of $J$ and for small geometric misalignments.

\subsection*{Coupling regimes}
\begin{figure*}[tb]
\centering
\includegraphics[width=0.8\textwidth]{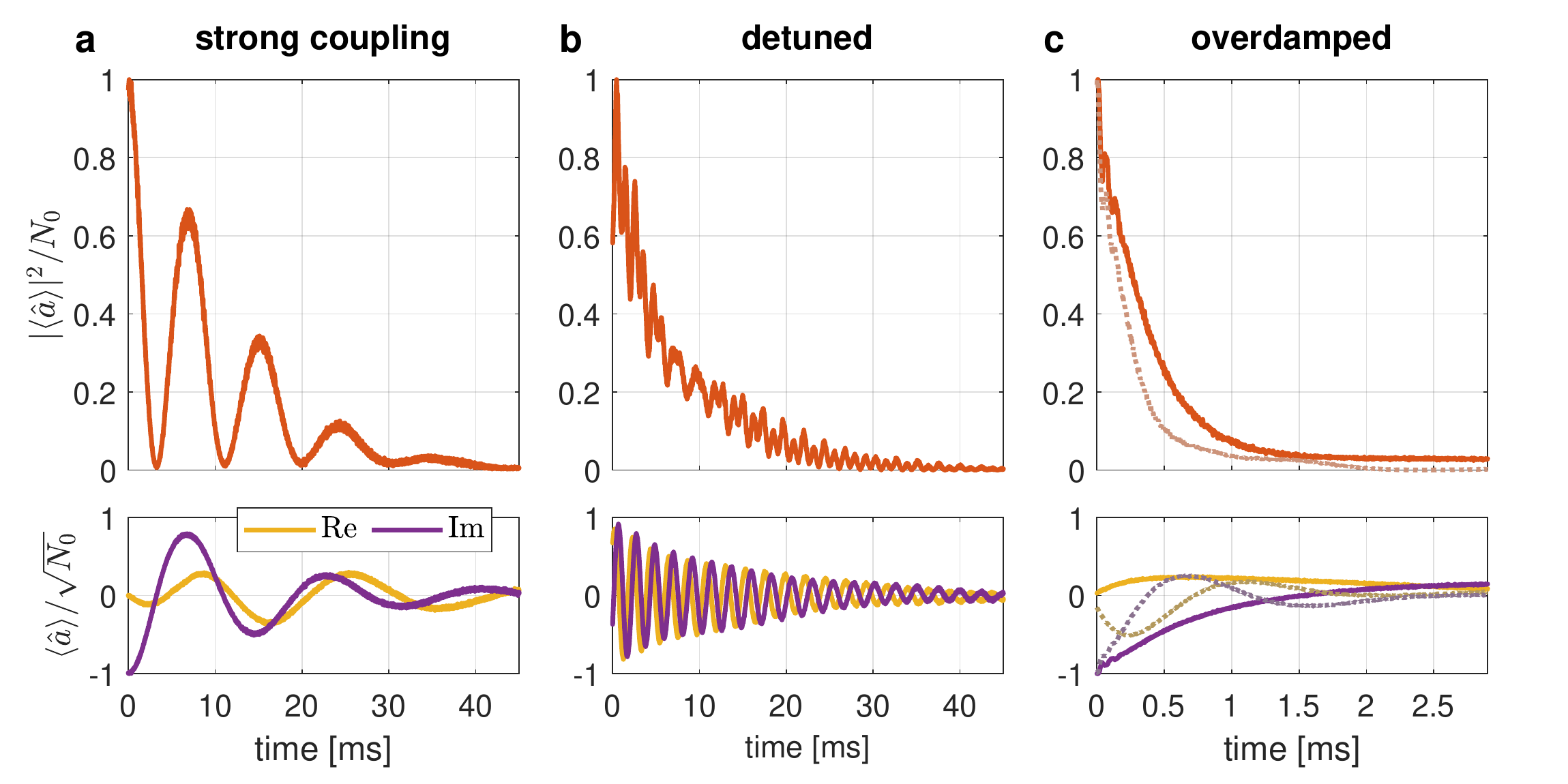}
\caption{\label{fig:temporal-and-regimes}
\textbf{Measured dynamics of the coupled alkali-metal--noble-gas spin system in three regimes}.
All measurements begin with a short magnetic stimulation of $N_0=(5.9\pm0.4)\cdot10^{13}$ alkali spin excitations, and the initial alkali-metal--noble-gas coupling rate is $J\approx 68$ \HzUnit{}.
\textbf{Top:} Collective spin excitations of the alkali atoms. \textbf{Bottom:} Real and imaginary parts of the collective spin amplitude, associated with the two transverse spin components in the lab frame, exhibiting Larmor precession in addition to the exchange.
\textbf{a.} \textbf{Strong-coupling}, achieved when $J$ exceeds the alkali relaxation rate $\gamma=0.11 J$ and close to resonance $\Delta= -1.15J$.
Recurring collapse and revival of alkali spin excitations provide evidence for a coherent hybridization with the noble-gas spins.
\textbf{b.} \textbf{Decoupled dynamics}, observed when increasing the detuning to $\Delta=6.8J=66\gamma$ by increasing the magnetic field.
The alkali spin, here largely decoupled from the noble-gas spin, undergoes standard Larmor precession and relaxation.
\textbf{c.} \textbf{Overdamped dynamics}, obtained at $\gamma=3.2J$. When near resonance (solid line, $\Delta=-0.15\gamma$), the long-lived noble-gas spin partially hybridizes with the alkali spin, whose relaxation slows down compared to the non-resonant case (dotted line, $\Delta=\gamma$).
}
\end{figure*}

It is instructive at this point to compare the resonant, strong-coupling dynamics to the detuned and overdamped dynamics.
These are presented in Fig.~\ref{fig:temporal-and-regimes}, showing the measured total number $|\langle\hat{a}\rangle|^2$ of collective alkali spin excitations (top panel) and the amplitudes Re$\langle\hat{a}\rangle$ and Im$\langle\hat{a}\rangle$, which exhibit also Larmor precession (bottom panel). 
For the coherent spin states in our experiment, oscillations of $|\langle\hat{a}(t)\rangle|^2$ correspond to nutations (tilt) of the collective alkali spin from the quantization axis $\hat{z}$ and therefore manifest the exchange interaction in a rotating frame, free of Larmor precession in the $xy$ plane.

First, we set $\Delta$ close to resonance ($\Delta=-1.15J$ as before) and measure the dynamics under the strong-coupling conditions $J=68\pm 5$ \HzUnit{} and $\gamma=7.5\pm2$ \HzUnit{} (Fig.~\ref{fig:temporal-and-regimes}a).
As in Fig.~\ref{fig:exchange}, we observe oscillations of the number of alkali spin excitations $|\langle\hat{a}(t)\rangle|^2$, exchanged back and forth with the noble-gas spin while gradually decaying.
The dynamics far-off resonance is shown for an increased detuning $\Delta=460$ \HzUnit{} $\approx6.8J$ (Fig.~\ref{fig:temporal-and-regimes}b). In this regime, we observe a decaying precession of $\langle\hat{a}(t)\rangle$ and an almost monotonic relaxation of $|\langle\hat{a}(t)\rangle|^2$ at a rate $14\pm2$ \HzUnit{}, in agreement with the expected value ($2\gamma$).
Finally, we repeat the experiments with an increased relaxation rate $\gamma=215$ \HzUnit{} $\approx 3.2J$ (Fig.~\ref{fig:temporal-and-regimes}c), implemented by keeping the pumping beam on during the measurement.
The measurements in Fig.~\ref{fig:temporal-and-regimes} of the three regimes elucidate the coherent nature of the exchange interaction under the strong-coupling conditions.

The transition between the overdamped and strong-coupling regimes is continuous, with the reversible dynamics becoming gradually more dominant.
At critical damping $J=\gamma/2$, the decay of the alkali spin is shared among both species, and its coherence time is effectively elongated.
Figure~\ref{fig:temporal-and-regimes}c demonstrates the elongation for on-resonance dynamics (solid) compared to the detuned case (dotted).
While $J/\gamma>0.5$ promotes an avoided crossing of the normal frequencies of the dynamics (as discussed below), the reversible exchange is negligible at critical damping.
At $J/\gamma=0.78$, for example, as realized by Kornack and Romalis \cite{KornackRomalis2002OverlappingEnsemles}, only $0.5\%$ of the initial excitations return to the initially tilted gas.
Efficient and reversible exchange of excitations, therefore, requires the ratio $J/\gamma$ to be large.

\subsection* {Spectral map}

\begin{figure*}[tb!]
\centering
\includegraphics[width=0.9\textwidth]{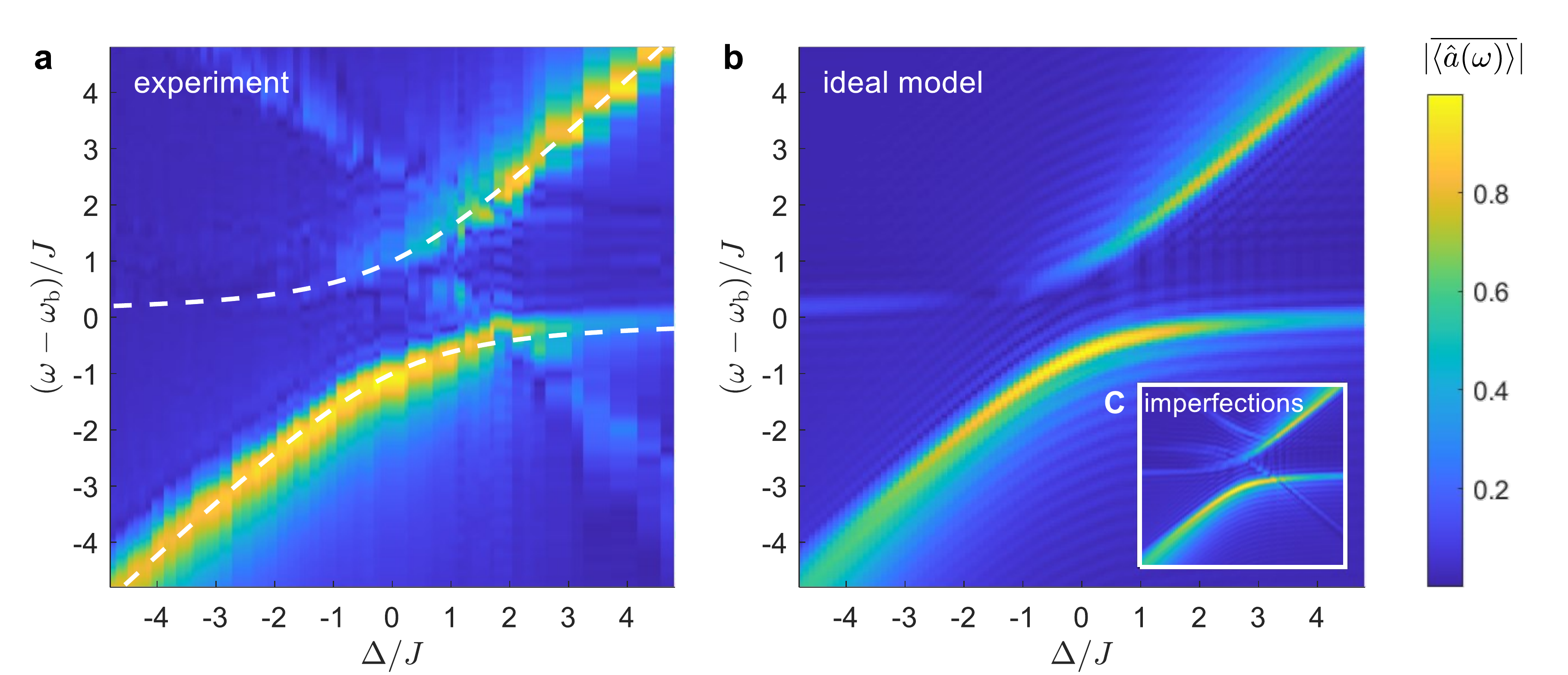}
\caption{
\label{fig:avoided-crossing} \textbf{Spectral response of the alkali-metal--noble-gas spin system in the strong coupling regime.}
\textbf{a.} Measured response of the collective alkali spin $\ea$ to a weak stimulation, for different detunings between the spins $\Delta$.
The spectrum $\overline{\langle \hat{a}(\omega) \rangle} \propto \int_0^\infty \langle \hat{a}(t) \rangle e^{-i\omega t} dt$ (normalized separately for each $\Delta$, see \Methods{}) manifests the eigenfrequencies of the coupled system;
The spectrum maxima correspond to the normal frequencies, and the spectral widths are indicative of the decay.
Dashed lines are the imaginary part of the eigenvalues of Eq.~(\ref{eq:coupled-spins}).
A clear avoided crossing with a sizeable spectral gap at $|\Delta|<J$ indicates the strong, coherent hybridization of the two spin gases.
The response at $|\Delta|>J$ corresponds to the independent precession rates of the alkali and noble-gas spins $\omega_\ra$ and $\wZ$ respectively.
The axes are scaled by the average value $J=47$ \HzUnit{} (rather than the initial value $J=78$ \HzUnit{}) to account for the decrease of $J$ due to alkali depolarization during the 65-ms-long measurement.
The frequency axis is shifted by $\wZ=42.9$ \HzUnit{}.
\textbf{b.} Calculated spectral response from a detailed model.
\textbf{c.} An experimental misalignment of 4.4 mrad between the magnetic field and the pumping direction, when added to the calculation, reproduces the weak perpendicular branches.
}
\end{figure*}

At strong coupling, the system's response to magnetic fields features a spectral gap.
We measure this gap by repeating the experiment presented in Fig.~\ref{fig:temporal-and-regimes}a for different values of $\Delta$.
The spectral map, shown in Fig.~\ref{fig:avoided-crossing}a, reveals an avoided crossing between the normal frequencies at $\Delta=0$; with a wide gap indicating a strong coherent coupling between the two gases.
We further compare the measurements to calculated spectra.
We present both a simple model based on Eq.~(\ref{eq:coupled-spins}) (dashed lines in Fig.~\ref{fig:avoided-crossing}a) and the results of the detailed model (Fig.~\ref{fig:avoided-crossing}b,c).
Both models reproduce well the main frequency branches.
The additional features in the spectrum, primarily the weak perpendicular branches and the vanishing amplitude of the horizontal branch at $\Delta\gtrsim J$ (due to reduced sensitivity to magnetic stimulation near the so-called compensation point \cite{KornackRomalis2002OverlappingEnsemles}) are well captured by the detailed model.

\section*{Discussion}

We realize strong coherent coupling between the collective spins of dense alkali-metal vapor and noble gas, with a coupling-to-decay ratio $J/\gamma\approx 10.7$ much larger than unity.
The coupling arises from accumulation of stochastic spin-exchange collisions, relying on the weakness of each collision (spin precession of $\sim10^{-5}$ radians per collision) to conserve coherence and reversibility \cite{weakcollisions2019arxiv}.
We estimate that higher values of $J/\gamma$ are achievable with higher $^3$He density and polarization and at lower temperature and nitrogen-gas pressure.
$^3$He pressure exceeding 10 atm was demonstrated \cite{Walker2017He3review} as well as 85\% polarization \cite{ChenBabcockWalker2014HigherLimitsSEOPhelium3}.
A system at $220\,^\circ$C with 8.2 atm of $^3$He polarized to 80\% and near-unity polarized potassium is expected to reach $J/\gamma>100$.

Operation of alkali and noble-gas systems in the strong-coupling regime opens several intriguing possibilities.
One route motivated by quantum information applications is using the alkali spins as mediators between photons and noble-gas spins \cite{noblegasSpectroscopy2020arxiv,noblegasStoragePRA2020arxiv}.
In particular, fast on-resonance coupling between the spins can enhance the indirect coupling to photons and improve the performance of these applications compared to detuned operation.
At strong coupling, read-in, read-out, and control of the collective noble-gas spin are done at a rate $J$, whereas detuned operations with $\Delta \gg J$ are done at a significantly lower rate $J^2/\Delta$ (up to Hz-scale).
The efficiency and fidelity of the operation are application-dependent and could be optimal in either of the two regimes.
For long-lived optical quantum memories, the optimal regime depends on the bandwidth $B$ of the optical signal \cite{noblegasStoragePRA2020arxiv}.
In systems with $J\gtrsim \gamma$, storage of photons with an optical bandwidth of $B\gg\gamma$ (kHz up to GHz) is optimized by first storing the light on the alkali spins and then transferring it to the noble-gas spins via a strong-coupling exchange.
The exchange efficiency, approximately $\mathrm{exp}[-\pi\gamma/2J]$, approaches unity for $J\gg\gamma$ and could enable hours-long storage with unprecedented time-bandwidth product.
Another example relates to the generation of long-lived spin-entanglement between multiple cells via detuned operation \cite{AlkaliNobleEntanglementKatz2020PRL}.
Once the entanglement is generated, efficient extraction for subsequent usage requires transfer to the alkali spins, which would rely on strongly-coupled exchange.
Furthermore, for $J\gg \gamma$, the generation is significantly more efficient regardless of the detuning, as the contribution of alkali projection noise is suppressed by a factor $4\gamma B/J^2$, where $B^{-1}$ is the duration of the entangling pulse.

A second potential route is utilizing the strong coupling for improving noble-gas-based sensors.
Noble-gas magnetometers sense magnetic fields by measuring the precession of noble-gas spins with an additional auxiliary magnetometer.
They are particularly interesting being fundamentally limited by the low noble-gas projection noise \cite{Walker2017He3review}.
Using the alkali spin as an embedded magnetometer enables to suppress its projection-noise contribution by a large factor over the projection noise of the noble gas when $J\gg\sqrt{\gamma/T_2^{\rb}}$ \cite{weakcollisions2019arxiv}.
In addition, operation in the strong coupling regime outperforms detuned operations in two aspects.
First, a detuned operation reduces the magnetic sensitivity at low frequencies near the NMR frequency of the noble-gas, effectively increasing the impact of photon shot noise.
Second, bringing the two spins to a resonance implies lowering the alkali precession frequency, which in turn enables operation in the spin-exchange relaxation-free regime where sensitivity is increased even further \cite{AppeltHapper1998SEOPtheoryPRA,Allred2002RomalisSERFmagnetometer}.
Finally, another application regards self-compensating magnetometers, which typically operate on resonance \cite{Brown2010RomalisCPTviolation,Bloch2019Axions}.
These sensors can readily benefit from an enhanced coupling rate, which could provide for higher bandwidth and dynamic range.

\section*{MATERIALS AND METHODS}

\subsection*{The Holstein-Primakoff transformation from spins to bosonic excitations}
The states of the alkali and noble-gas spin ensembles are characterized by their degree of polarization $p_\ra=(2/N_\ra) \langle \sum_m \hat{\mathrm{s}}_z^{(m)} \rangle$ and $p_\rb=(2/N_\rb) \langle \sum_n \hat{\mathrm{k}}_z^{(n)} \rangle$.
Here, $\sum_{m}\hat{\mathrm{s}}^{(m)}_{j}$ and $\sum_{n}\hat{\mathrm{k}}^{(n)}_{j}$ with  $j=\{x,y,z,-,+\}$ are the standard collective spin operators of the electrons of the alkali atoms and the nuclei of the noble gas atoms, respectively, and $N_\ra = n_\ra V$ and $N_\rb = n_\rb V$ are the number of atoms in the volume $V$.
Describing the alkali spins in terms of only the electronic spins is possible owing to the frequent alkali-alkali collisions, which constantly drive the alkali atoms to a spin-temperature distribution \cite{Walker1997SEOPReview}.
In the spin-temperature distribution, due to the hyperfine coupling to the alkali nuclear spin, the spin precession around an external magnetic field is slower than that of a bare electron by a factor $q(p_\ra)$, known as the slowing-down factor; for potassium, $q(p_\ra) = 2+4/(1+p_\ra^2)$ \cite{Allred2002RomalisSERFmagnetometer,Vasilakis2011RomalisBackactionEvation,Walker1997SEOPReview}.

We are interested in the bosonic annihilation operators $\ha$ and $\hb$, defined according to the Holstein-Primakoff transformation as $\hat{a}=\sqrt{q/N_\ra p_\ra} \sum_{m}\hat{\mathrm{s}}^{(m)}_{-}$ and $\hat{b}=\sqrt{1/N_\rb p_\rb }\sum_{n}\hat{\mathrm{k}}^{(n)}_{-}$ \cite{Polzik2010ReviewRMP,weakcollisions2019arxiv}.
These are the canonical, normalized version of the collective spin operators transverse to the quantization axis.
For the alkali spins, we denote the homogeneous depolarization rate by $\Gamma_\rp=-(\partial_t p_\ra)/ p_\ra$ and the transverse relaxation rate by $\Gamma_2=\Gamma_\rc+\Gamma_\rp$ (decay rate of $\langle \sum_m \hat{\mathrm{s}}_x^{(m)}\rangle$ and $\langle \sum_m \hat{\mathrm{s}}_y^{(m)}\rangle$), where $\Gamma_\rc$ is the dephasing rate. The decoherence rate of the excitations $\langle \hat{a}\rangle$ is therefore given by $\gamma=\Gamma_\rc+\Gamma_\rp/2$, neglecting small variations of $q$ on short timescales.
The noise on  $\langle \hat{a}\rangle$ (technical or fundamental) signifies incoherent excitations, which inevitably increase when the polarization decays.
As a result, the process of depolarization ($\Gamma_\rp>0$), while contributing only partially ($\Gamma_\rp/2$) to the collective spin decoherence, contributes as well to the increase of the fundamental and technical noises.

\subsection*{Apparatus and experimental conditions}
We use a spherical cell with diameter $\ell=2.54$ cm and volume $V=\unit[8.6]{cm^3}$, made of GE-180 aluminosilicate glass, containing $^3$He gas, a droplet of natural abundant potassium, and 50 Torr of nitrogen.
The temperature of the cell $T=\unit[230]{^{\circ}C}$ is maintained using a pair of resistance twisted wires wrapped around an alumina body, which are driven with current oscillating at 320 kHz.
The magnetic field is applied via three sets of coils: 4-winding double Helmholtz coils for controlling $B\hat{z}$ and a bird-cage coil for the transverse fields to improve magnetic uniformity.
The coils are placed inside five concentric layers of $\mu$-metal magnetic shields, and the inner two layers are degaussed.

The $N_\ra=4.2\cdot10^{15}$ potassium atoms are polarized by optical pumping using 500 mW of circularly-polarized light at 770 nm.
This pumping light is generated using a free-running diode laser followed by a tapered amplifier.
We tune the laser near the optical D1 transition, which in our setup appears as a single absorption line with a full width of 32 GHz due to pressure broadening, producing an on-resonance optical depth of $n_\ra \sigma_\text{abs} \ell\approx$ 220 ($\sigma_\text{abs}=1.76\cdot10^{-13}$ cm$^2$ is the absorption cross-section of the 32-GHz-wide line).
The pumping beam is Gaussian with a 25-mm waist diameter. We detune it from resonance to reduce its depletion and achieve the high degree of spin polarization $p_\ra \ge 0.95$.

Spin-destruction collisions among potassium atoms and spin-rotation interaction of potassium atoms with the buffer gas dominate the depolarization of the potassium spins in the dark \cite{Happer2010book}.
The depolarization is generally a multi-exponential process (as can be seen in Fig.~\ref{fig:PandQ}b), and yet, at short times, it can be described by the single rate $\Gamma_\rp=11.4$ \HzUnit{}.
Rapid spin-exchange collisions among potassium atoms at a rate $R_{\mathrm{se}}=86$ kHz and the operation at low Larmor precession rates $|\omega_\ra| \ll \sqrt{ R_\mathrm{se} \Gamma_\rp }$ puts the potassium in the so-called spin-exchange relaxation-free (SERF) regime \cite{Happer1977SERF,Allred2002RomalisSERFmagnetometer}, rendering the relaxation induced by spin-exchange collisions negligible.
Consequently, the transverse spin relaxation rate $\Gamma_2=\Gamma_\rp+\Gamma_\rc=13$ \HzUnit{} is dominated by the depolarization processes, with minor contribution from magnetic inhomogeneity ($\Gamma_\rc=1.6$ \HzUnit{}).
These lead to a decoherence rate of $\gamma=\Gamma_\rc+\Gamma_\rp/2\approx 7.3$ \HzUnit{} for the bosonic excitations of the collective potassium spin.

\begin{figure}[tb]
\centering
\includegraphics[width=1.0\columnwidth]{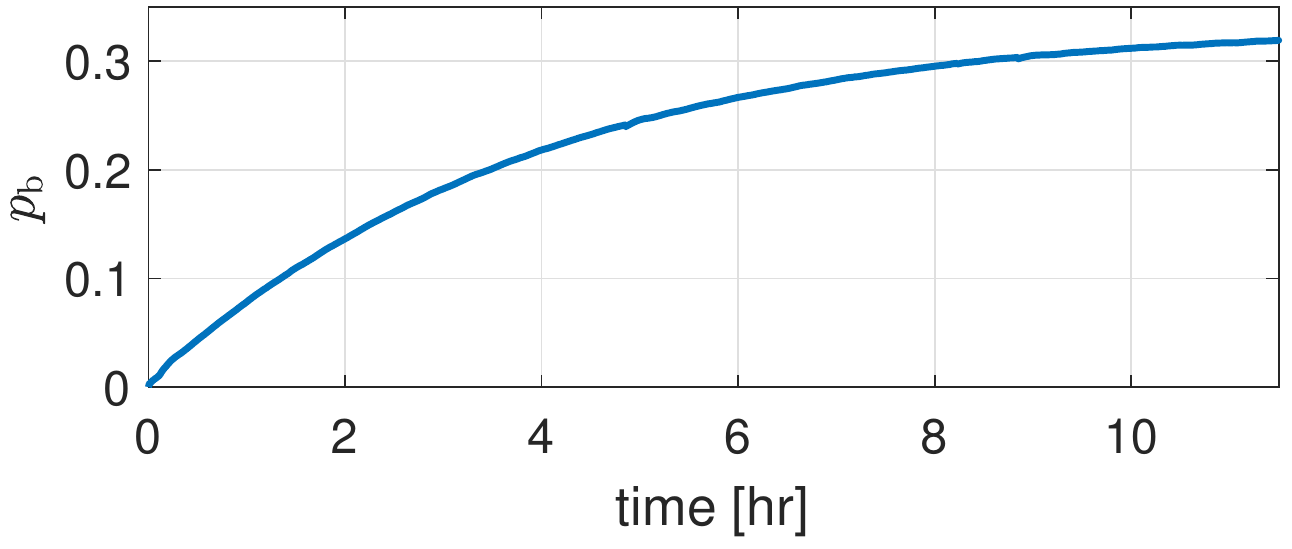}
\caption{\label{fig:SEOP}
\textbf{Spin-exchange optical pumping.}
Typical measurement of the pumping process of helium-3 by optically pumped potassium vapor.
Here the potassium density is $n_\ra=4.9\cdot10^{14}~/\mathrm{cm}^3$, and the helium depolarization time is $T^\rb_{1,\mathrm{act}}=3.9$ hours.
}
\end{figure}

The $N_\rb=5.5\cdot10^{20}$ helium atoms are hyperpolarized using spin-exchange optical pumping (SEOP) \cite{Walker1997SEOPReview} at a rate $3.6\cdot 10^{-6}$ \HzUnit{} in the presence of an axial magnetic field $B=400$ mG.
A typical SEOP measurement settling at $p_\rb\ge 0.3$ is presented in Fig.\,\ref{fig:SEOP}.
In our system, at low temperature, the measured depolarization and decoherence times of the helium spins  $T^\rb_1=22$ hours and $T^\rb_2=2$ hours are limited by magnetic field inhomogeneity within the cell volume.
At elevated temperature and polarizations, we measure $T^\rb_{1,\mathrm{act}}=3.9$ hours (see Fig.~\ref{fig:SEOP}) due to inhomogeneity of the magnetizations of the two ensembles in the cell, which slightly deviates from an ideal sphere \cite{Romalis2014CommentGradientsSphere}.
To moderate the helium depolarization during the experiments, we intermittently apply SEOP conditions in-between measurements.

The polarized spin ensembles exert an equivalent magnetic field (EMF) on each other, via collisions and via the macroscopic magnetic fields generated by their magnetization.
While the EMF experienced by the helium $B_{\ra\rightarrow\rb}=-0.24$ mG (for $p_\ra= 0.98$) is small, the EMF experienced by the potassium $B_{\rb\rightarrow\ra}=-10.94$ mG (for $p_\rb = 0.3$) is considerable.
The detuning from resonant coupling $\Delta$ is thus quite sensitive to $p_\rb$, which we monitor during the experiment.
We do so by applying a constant magnetic field $-B_{\rb\rightarrow\ra}+1.6$ mG, and monitoring the precession frequency of the decoupled alkali spins following a small transverse magnetic pulse.

In the experiments presented in Fig.~\ref{fig:exchange}, Fig.~\ref{fig:temporal-and-regimes}, and Fig.~\ref{fig:avoided-crossing}, we use a transverse magnetic pulse to tilt the alkali spins by $\theta_\ra=9.8\pm 0.2^\circ$,  $\theta_\ra=6.8\pm 0.2^\circ$, and $\theta_\ra=0.7\pm 0.05^\circ$, respectively.
In terms of the number of excitations $|\langle \hat{a} \rangle|^2=q N_\ra p_\ra \theta_\ra^2 /4$, these correspond to $|\langle \hat{a} \rangle|^2=(12.1\pm0.5)\cdot10^{13}$, $|\langle \hat{a} \rangle|^2=(5.9\pm0.4)\cdot10^{13}$, and $|\langle \hat{a} \rangle|^2=(6.2\pm0.9)\cdot10^{11}$.

In all experiments, we measure the transverse spin component of the alkali atoms along the $\hat{x}$ axis using Faraday rotation of a linearly-polarized probe beam.
The 5-mm diameter, 260-$\mu$W probe beam is detuned by ${\sim}400$ GHz  above the D1 transition, and its polarization is measured after the cell using balanced photodetection method \cite{Katz2015SERFHybridization}.
We subtract from all measurements a background signal taken without the magnetic pulse.
This background signal is small and is dominated by excitations of transverse spins during the fast variation of $B\hat{z}$ (when setting $\Delta$), due to imperfect alignment between the optical and magnetic axes.

\subsection*{Reconstruction and scaling of $\langle\hat{a}\rangle$ and $\langle\hat{b}\rangle$}

We use optical Faraday rotation to measure $\langle\hat{a}\rangle$ and $\langle\hat{b}\rangle$.
For the optically-broadened line and the far-detuned probe in our setup, and as long as the Faraday-rotation angle is small, the balanced-detection readout is proportional to the $\hat{x}$ component of the collective alkali spin $\langle\sum_m\hat{\mathrm{s}}^{(m)}_x\rangle$, {\it i.e.}, to the electron spin projection along the probing axis \cite{Duan2000FaradayRotation}.
From these measurements we extract the normalized transverse spin component $\bar{S}_x(t)=\langle\sum_m\hat{\mathrm{s}}^{(m)}_x(t)\rangle/[N_\ra p_\ra(0)/2]$.
The normalization factor is calibrated separately by tilting the initial spin $[N_\ra p_\ra(0)/2]\hat{z}$ all the way to the $\hat{x}$ direction (equivalent to $\theta_\ra=90^\circ$) and measuring the maximal Faraday rotation angle ($\sim$4 radian in our system).
We verify that the Faraday rotation angle in all subsequent experiments is small.

\begin{figure}[tb]
\centering
\includegraphics[width=1.0\columnwidth]{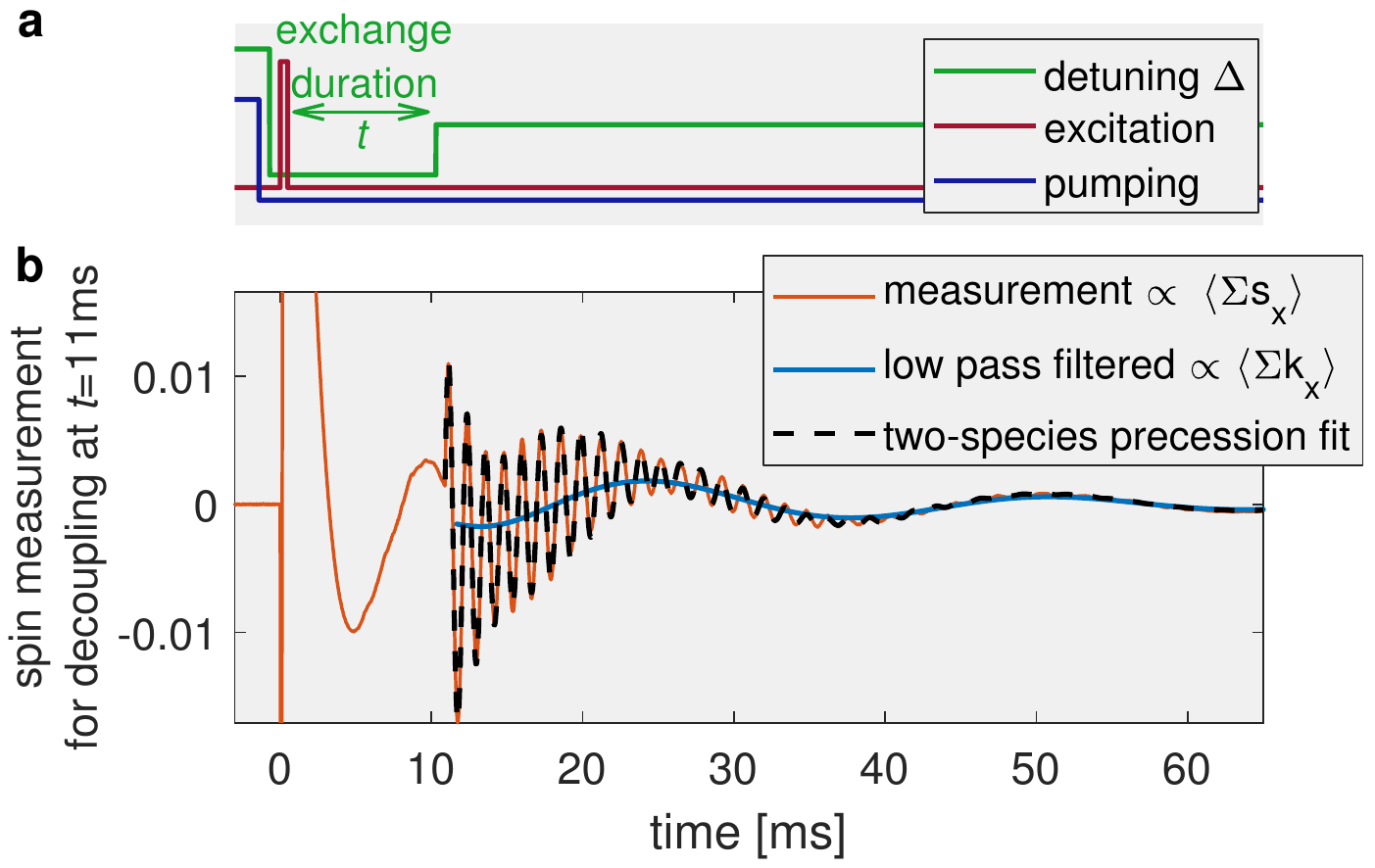}
\caption{\label{fig:sequence}
\textbf{Pulse sequence and typical results of an excitation-exchange measurement.}
\textbf{a.} First, we turn off the pumping and bring the two species to strong coupling with a small detuning $\Delta$.
We then generate a transverse excitation with a pulse of transverse magnetic field.
At a later time $t$, we halt the exchange by increasing the axial magnetic field and setting a large $\Delta$.
\textbf{b.} Example of a measured signal with exchange duration $t=11$ ms, with $\Delta=-1.15J$ before $t$, and $\Delta=790$ Hz $\gg J$ after $t$.
We measure the alkali electron spin (red) which, once $\Delta$ is increased, can be used as a magnetometer that senses the noble-gas spin. The fast oscillations of the signal correspond to the Larmor precession of the alkali spin, and the slow modulation corresponds to the noble-gas precession.
The latter is highlighted by the blue line (generated by low-pass filtering of the signal for illustrative purposes).
We fit the signal to the model from Eq.~(\ref{eq:fitted-model}) (dashed black line) and find the amplitudes of the alkali and noble-gas components at time t, which are used to estimate $\langle \hat{a}(t) \rangle$ and $\langle \hat{b}(t) \rangle$, respectively. The same fit also provides $p_\mathrm{a}(t)$.
}
\end{figure}

The measurements of $|\langle \hat{a}(t) \rangle|^2$ and $|\langle \hat{b}(t) \rangle|^2$ presented in Fig.~\ref{fig:exchange} are done according to the experimental sequence shown in Fig.~\ref{fig:sequence}a).
The sequence starts by initializing the spins with a small transverse component under conditions of small $\Delta$.
After some evolution and partial decay in the dark, at time $t$, we increase $\Delta$ by an order of magnitude (by increasing $B+B_{\rb\rightarrow\ra}$ to $1.5$ mG), thus largely decoupling the alkali and noble-gas spins.
We continue to monitor the alkali spins and use them as a magnetometer for sensing the noble-gas spins.

\begin{figure}[tb]
\centering
\includegraphics[width=1.0\columnwidth]{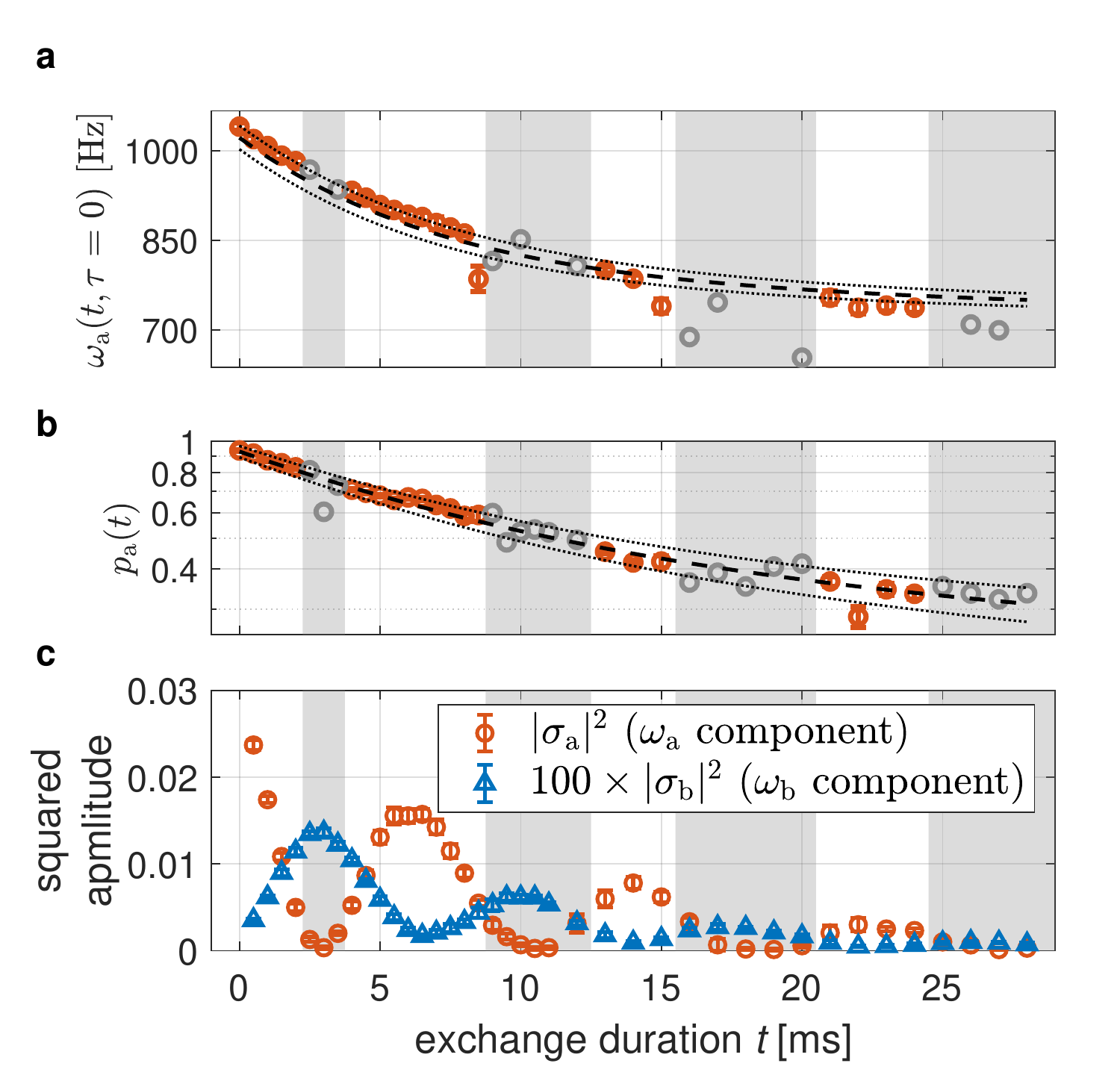}
\caption{
\textbf{Variables extracted from fitting Eq.~(\ref{eq:fitted-model}) to the measured signals for each exchange time $t$, as exemplified in Fig.~\ref{fig:sequence}.
a.} The change in alkali precession frequency $\omega_a(t,\tau=0)$ [see Eq.~(\ref{eq:omega-a})] manifests the change in the slowing-down factor due to alkali depolarization.
\textbf{b.} The degree of alkali polarization $p_\ra(t)$ (in semi-log scale).
In \textbf{a} and \textbf{b}, dashed black line correspond to the fitted multi-exponential model, and dotted lines present its confidence bounds.
These are used as uncertainty estimations when using $p_\mathrm{a}(t)$ to scale $\langle \hat{a} \rangle$ and $\langle \hat{b} \rangle$. Less reliable data, extracted when the excitations reside predominantly in the noble gas spins, are marked in gray.
\textbf{c.} The two frequency components (amplitude squared) of the normalized Faraday rotation signal $\bar{S}_x(t+\tau)$. Note the factor of $(\Delta/J)^2 \gtrsim 100$ between them.
Each data-point is averaged over 12 to 20 repetitions of the sequence
}
\label{fig:PandQ}
\end{figure}

During the experiment, when the pumping light is off, the polarization of the alkali spin decays $p_\ra(t)\le p_\ra(0)$.
This decay changes the slowing-down factor $q(t)=q[p_\ra(t)]$ and thus shifts the Larmor precession frequency of the alkali spin, which we directly measure (see Fig.~\ref{fig:PandQ}a).
We model the time dependence of the shift assuming an exponential polarization decay $\partial_\tau p_\ra = -\tilde{\Gamma}_\rp p_\ra$, where $\tau$ is the time elapsed from the decoupling time $t$, and $\tilde{\Gamma}_\rp$ is the depolarization rate at the increased magnetic field.
The instantaneous precession frequency of the alkali spin is then given by
\begin{equation} \omega_\ra(\omega_0,p_\ra(t),\tilde{\Gamma}_\rp;\tau) = \frac{2\omega_0}{ 1+2/[1+p_\ra^2(t)e^{-2\tilde{\Gamma}_\rp \tau}]},\label{eq:omega-a}
\end{equation}
where $\omega_0 = \omega_\ra(p_\ra=1;\tau=0)$.
To each measured signal $\bar{S}_x$, we therefore fit the model
\begin{align}
\bar{S}_x(t+\tau) & = \mathrm{Re} \left[\sigma_\ra(t) e^{ i\int_0^\tau \omega_\ra(\omega_0,p_\ra(t),\tilde{\Gamma}_\rp;\tau^\prime) d\tau^\prime - \gamma_\ra \tau } \right. \nonumber \\
& \left. + \sigma_\rb(t) e^{ ( i\wZ - \gamma_\rb ) \tau } \right].
\label{eq:fitted-model} 
\end{align}
Here $\sigma_\ra(t)$, $\sigma_\rb(t)$ are complex fitting parameters, corresponding to the amplitudes of the two frequency components, and $\gamma_\ra$, $\gamma_\rb$, $\omega_0$, $\wZ$, $p_\ra(t)$ are real fitting parameters.
One such fit is demonstrated in Fig.~\ref{fig:sequence}b, and the extracted $p_\ra(t)$, $\omega_\ra(t,\tau=0)$, $|\sigma_\ra(t)|^2$, and $|\sigma_\rb(t)|^2$ are shown in Fig.~\ref{fig:PandQ}c) [note the factor $| \sigma_\rb / \sigma_\ra |^2 \approx (J/\Delta)^2 < 1/100$].
We verify that the extracted amplitudes are insensitive to the exact value of $\tilde{\Gamma}_\rp$ (set to be 8.6 \HzUnit{} in all fits) and even to the functional form of $p_\ra(t+\tau)$.
We use the fits to estimate $p_\ra(t)$ (Fig.~\ref{fig:PandQ}b) and find that it is accurately described by the double-exponent function  $p_\ra(t)=0.61e^{-t/(9.1~\mathrm{ms})}+0.381e^{-t/(102~\mathrm{ms})}$, presented in Fig.~\ref{fig:PandQ}b) with its confidence bounds.
The multi-exponential nature of the depolarization can be attributed to multi-mode spatial dynamics \cite{Shaham2020Diffusion}, to SEOP of the alkali by the noble gas, and to low signal-to-noise ratios.

With $p_\ra(t)$ at hand, we obtain the factor $\eta(t)= \frac{N_\ra q[p_\ra(t)] p_\ra^2(0)}{4  p_\ra(t)} $ between the number of alkali excitations and $\bar{S}_x^2$.
The alkali and noble-gas excitations presented in Fig.~\ref{fig:exchange} are then given by $|\langle\hat{a}(t)\rangle|^2= \eta(t)\big|\sigma_\ra(t) + \sigma_\rb(t)\big|^2$ and $|\langle\hat{b}(t)\rangle|^2 = \eta(t)\big| \frac{\Delta(t)}{J(t)}\sigma_\rb(t) -\frac{J(t)}{\Delta(t)}\sigma_\ra(t)\big|^2$, where $\Delta(t) = \omega_\ra(t,\tau=0) - \wZ$ and $J(t)=\sqrt{ \frac{p_\ra(t)}{p_\ra(0)} \frac{q(0)}{q(t)} } J(t=0)$.
These expressions neglect terms of order $(J/\Delta)^2$ and higher.
Here we see that the decay of $p_\ra(t)$ increases $\eta$, justifying the relation $\gamma \le  \Gamma_2$ and explaining the growth of noise in the measurements, which one can interpret as excess thermal excitations.
This process leads to the increasing errorbars in Fig.~\ref{fig:exchange} at later times.
For the experiments presented in Figs.~\ref{fig:temporal-and-regimes} and \ref{fig:avoided-crossing}, we reconstruct the complex-valued $\langle\hat{a}(t)\rangle = \sqrt{\eta(t) } [\bar{S}_x(t)-i \bar{S}_y(t)]$.
The two normalized projections $\bar{S}_x(t)$ and $\bar{S}_y(t)$ are measured in two consecutive experiments that differ in the direction of the initial pulsed excitation (alternating between $B_\perp\hat{y}$ and $B_\perp\hat{x}$).
In Fig.~\ref{fig:exchange}, we average the measurements with the two excitation directions.
We use the extracted $p_\ra(t)$ for Figs.~\ref{fig:temporal-and-regimes}a and \ref{fig:temporal-and-regimes}b and estimate $p_\ra=p_\ra(t=0)=0.98$ for Fig.~\ref{fig:temporal-and-regimes}c.
Finally, in Fig.~\ref{fig:avoided-crossing} we present the normalized Fourier amplitudes $|\overline{\langle\hat{a}(\omega)\rangle}| = |\int_0^\infty \langle \hat{a}(t) e^{-i\omega t} \rangle dt| /\sqrt{T \int_0^\infty |\langle \hat{a}(t)\rangle|^2 dt}$, where $T=65$ ms is the sequence duration.

\subsection*{Detailed model}

Equation (\ref{eq:coupled-spins}) describes the idealized dynamics of the spin gases.
For the calculations presented in Figs.~\ref{fig:exchange} and \ref{fig:avoided-crossing}b,c, we use a detailed model, which includes the decay of the alkali polarization $p_\ra=p_\ra(t)$ during the experimental sequence, the dependence of $\Delta$ on $p_\ra(t)$ via the slowing-down factor $q[p_\ra(t)]$, misalignment of the optical and magnetic axes, and residual transverse magnetic fields.

The model assumes that both spin ensembles are initially polarized along $-\hat{z}$.
It follows Refs.~\cite{Walker1997SEOPReview,weakcollisions2019arxiv} and describes the dynamics of the collective spin excitations $S_-= \langle \sum_m \hat{s}_-^{(m)} \rangle$ and $K_-= \langle \sum_n \hat{k}_-^{(n)} \rangle$, coupled by the Fermi-contact interaction occurring during stochastic collisions.
In the presence of axial magnetic field $B\hat{z}$ and transverse magnetic field $B_- = B_x -i B_y$, the coupled spin equations are given by
\begin{align} 
\pt S_- & =
i (\omega_\ra+i \Gamma_2)S_- -i \frac{n_\ra}{q n_\rb} J_\ra K_- +i \frac{g_e}{q}\frac{N_\ra p_\ra}{2} B_- , \nonumber \\
\pt K_- & = -i \frac{q n_\ra}{n_\rb} J_\rb S_- +i \wZ K_- +i g_\rb \frac{N_\rb p_\rb}{2} B_- . \label{eq:full-model}
\end{align}
Here $J_\ra=\sqrt{q} \tz n_\rb p_\ra /2$ and $J_\rb= \tz n_\ra p_\rb /2\sqrt{q}$ are the uni-directional coupling rates, eventually composing the bi-directional rate $J=\sqrt{J_\ra J_\rb}$, with $\tz = (\unitfrac[2\cdot10^{-14}]{cm^3}{s})/\sqrt{q}$.
The gyromagentic ratios of the electron and helium-3 spins are $g_e=-2.8\cdot10^6$ \HzUnit{}/G and $g_\rb=-3.24\cdot10^3$ \HzUnit{}/G, and the precession frequencies are $\omega_\ra = g_e B / q + \tz n_\rb p_\rb /2\sqrt{q} $ and $\wZ = g_\rb B + \sqrt{q}\tz n_\ra p_\ra /2$.

We simulate the experimental sequences by numerically solving these equations.
From the simulation results, we calculate the expectation values $\ea = \sqrt{q[p_\ra(t)]/N_\ra p_\ra(t)} S_-(t) $ and $\eb = \sqrt{1/N_\rb p_\rb} K_-(t) $.
For the model parameters, we use known constants or the measured values from the calibration experiments;
we use $p_\ra(t)=0.61e^{-t/(9.1~\mathrm{ms})}+0.381e^{-t/(102~\mathrm{ms})}$ for the alkali polarization and  $p_\rb=n_\ra k_\text{se} p_\ra(0) T^\rb_{1,\mathrm{act}}=0.32$ for the noble-gas polarization, where $k_\text{se}=5.5\cdot10^{-20}~\mathrm{cm^3/s}$ is the SEOP rate.

The model can account for various geometric misalignments and other experimental imperfections:
(1) Misalignment of the pumping beam from the $\hat{z}$ axis generates an initial transverse spin component.
If the pumping beam points towards $\eta_x\ex + \eta_y\ey+\ez$ (given $\eta_{x,y}\ll 1$), the initial value of $S_-$ is $N_\ra p_\ra(t=0) (\eta_x -i \eta_y) /2$.
(2) A residual magnetic field  pointing towards $\beta_x\ex + \beta_y \ey + \ez$ (given $\beta_{x,y}\ll 1$) during the SEOP process would turn the initial value of $K_-$ to $N_\rb p_\rb (\beta_x -i \beta_y) /2$.
(3) A non-vanishing transverse magnetic field during the sequences can be accounted for by a constant offset of $B_-$.
When varying $\Delta$ during the sequence, these misalignments could  tilt the spins and introduce spurious (background) excitations.
(4) A misalignment of the probe field can be accounted for by extracting the signal $S=\mathrm{Re}[(1+i\varepsilon_\Vert)S_-] +\varepsilon_\perp N_\ra p_\ra /2$ for $\varepsilon_{\Vert,\perp}\ll1$ (rather than simply $S=S_-$) from the simulation results.

We calculate the spectral map presented in Fig.~\ref{fig:avoided-crossing}b by repeating the calculation for $\unit[10.9]{mG}<B<\unit[11.8]{mG}$ (corresponding to $-5<\Delta/J<5$ for $J=47$ \HzUnit{} as used for normalizing Fig.~\ref{fig:avoided-crossing}).
The excitation is simulated by applying $B_\perp(t) = \unit[2.4]{mG} \times \exp[-t^2/(\unit[2.8]{\mu s})^2]$ (tilting by $\theta_\ra = 3^\circ$).
We calculate the Fourier transform of $\ea$ and the normalized amplitude $|\overline{ \langle \hat{a} (\omega) \rangle }|$, as done for the experimental data.
In Fig.~\ref{fig:avoided-crossing}b, we consider a perfectly aligned setup ($\beta_{x,y}=\eta_{x,y}=\epsilon_\Vert=\epsilon_\perp=0$).
In Fig.~\ref{fig:avoided-crossing}c, we reproduce the imperfection generating the perpendicular frequency branch by introducing a minute misalignment $\beta_x=3.1$~mrad and $\beta_y=-\beta_x$.
The calculations for Fig.~\ref{fig:exchange}a (solid lines) are done with $B=\unit[11.33]{mG}$ ($\Delta=-1.15J$) and $B_-(t) = \unit[5.2]{mG} \times \exp[-t^2/(\unit[2.8]{\mu s})^2]$ (tilting by $\theta_\ra=6.5^\circ$).

\section*{Acknowledgments:} 
We thank Chen Avinadav and Ran Finkelstein for fruitful discussions. We acknowledge financial support by the Israel Science Foundation, the European Research Council starting investigator grant Q-PHOTONICS 678674, the Minerva Foundation with funding from the Federal German Ministry for Education and Research, and the Laboratory in Memory of Leon and Blacky Broder.
\textbf{Author Contributions Statement:} All authors contributed to the experimental design, construction, data collection, and analysis of this experiment.
R.S. claims responsibility for all figures.
The authors wrote the manuscript together.
\textbf{Competing interests:} All authors declare that they have no competing interests.
\textbf{Data and materials availability:} All data needed to evaluate the conclusions in the paper are present in the paper.
The data that support the findings of this study and additional data are available from the corresponding author upon request.

\bibliography{strong_coupling}

\begin{thebibliography}{44}
\expandafter\ifx\csname natexlab\endcsname\relax\def\natexlab#1{#1}\fi
\expandafter\ifx\csname bibnamefont\endcsname\relax
  \def\bibnamefont#1{#1}\fi
\expandafter\ifx\csname bibfnamefont\endcsname\relax
  \def\bibfnamefont#1{#1}\fi
\expandafter\ifx\csname citenamefont\endcsname\relax
  \def\citenamefont#1{#1}\fi
\expandafter\ifx\csname url\endcsname\relax
  \def\url#1{\texttt{#1}}\fi
\expandafter\ifx\csname urlprefix\endcsname\relax\def\urlprefix{URL }\fi
\providecommand{\bibinfo}[2]{#2}
\providecommand{\eprint}[2][]{\url{#2}}

\bibitem[{\citenamefont{Gentile et~al.}(2017)\citenamefont{Gentile, Nacher,
  Saam, and Walker}}]{Walker2017He3review}
\bibinfo{author}{\bibfnamefont{T.~R.} \bibnamefont{Gentile}},
  \bibinfo{author}{\bibfnamefont{P.~J.} \bibnamefont{Nacher}},
  \bibinfo{author}{\bibfnamefont{B.}~\bibnamefont{Saam}}, \bibnamefont{and}
  \bibinfo{author}{\bibfnamefont{T.~G.} \bibnamefont{Walker}},
  \bibinfo{journal}{Reviews of Modern Physics} \textbf{\bibinfo{volume}{89}},
  \bibinfo{eid}{045004} (\bibinfo{year}{2017}).

\bibitem[{\citenamefont{Walker and Larsen}(2016)}]{WalkerLarsen2016NGCNMRG}
\bibinfo{author}{\bibfnamefont{T.~G.} \bibnamefont{Walker}} \bibnamefont{and}
  \bibinfo{author}{\bibfnamefont{M.~S.} \bibnamefont{Larsen}},
  \bibinfo{journal}{Advances in Atomic Molecular and Optical Physics}
  \textbf{\bibinfo{volume}{65}}, \bibinfo{pages}{373} (\bibinfo{year}{2016}).

\bibitem[{\citenamefont{Kornack and
  Romalis}(2002)}]{KornackRomalis2002OverlappingEnsemles}
\bibinfo{author}{\bibfnamefont{T.~W.} \bibnamefont{Kornack}} \bibnamefont{and}
  \bibinfo{author}{\bibfnamefont{M.~V.} \bibnamefont{Romalis}},
  \bibinfo{journal}{Physical Review Letters} \textbf{\bibinfo{volume}{89}},
  \bibinfo{pages}{253002} (\bibinfo{year}{2002}).

\bibitem[{\citenamefont{Jimenez-Martinez
  et~al.}(2014)\citenamefont{Jimenez-Martinez, Kennedy, Rosenbluh, Donley,
  Knappe, Seltzer, Ring, Bajaj, and
  Kitching}}]{JimenezMartinez2014KitchingXeNMR}
\bibinfo{author}{\bibfnamefont{R.}~\bibnamefont{Jimenez-Martinez}},
  \bibinfo{author}{\bibfnamefont{D.~J.} \bibnamefont{Kennedy}},
  \bibinfo{author}{\bibfnamefont{M.}~\bibnamefont{Rosenbluh}},
  \bibinfo{author}{\bibfnamefont{E.~A.} \bibnamefont{Donley}},
  \bibinfo{author}{\bibfnamefont{S.}~\bibnamefont{Knappe}},
  \bibinfo{author}{\bibfnamefont{S.~J.} \bibnamefont{Seltzer}},
  \bibinfo{author}{\bibfnamefont{H.~L.} \bibnamefont{Ring}},
  \bibinfo{author}{\bibfnamefont{V.~S.} \bibnamefont{Bajaj}}, \bibnamefont{and}
  \bibinfo{author}{\bibfnamefont{J.}~\bibnamefont{Kitching}},
  \bibinfo{journal}{Nature Communications} \textbf{\bibinfo{volume}{5}},
  \bibinfo{eid}{3908} (\bibinfo{year}{2014}).

\bibitem[{\citenamefont{Katz et~al.}(2020{\natexlab{a}})\citenamefont{Katz,
  Shaham, Polzik, and Firstenberg}}]{AlkaliNobleEntanglementKatz2020PRL}
\bibinfo{author}{\bibfnamefont{O.}~\bibnamefont{Katz}},
  \bibinfo{author}{\bibfnamefont{R.}~\bibnamefont{Shaham}},
  \bibinfo{author}{\bibfnamefont{E.~S.} \bibnamefont{Polzik}},
  \bibnamefont{and}
  \bibinfo{author}{\bibfnamefont{O.}~\bibnamefont{Firstenberg}},
  \bibinfo{journal}{Physical Review Letters} \textbf{\bibinfo{volume}{124}},
  \bibinfo{eid}{043602} (\bibinfo{year}{2020}{\natexlab{a}}).

\bibitem[{\citenamefont{Katz et~al.}(2020{\natexlab{b}})\citenamefont{Katz,
  Reches, Shaham, Gorshkov, and Firstenberg}}]{noblegasStoragePRL2020arxiv}
\bibinfo{author}{\bibfnamefont{O.}~\bibnamefont{Katz}},
  \bibinfo{author}{\bibfnamefont{E.}~\bibnamefont{Reches}},
  \bibinfo{author}{\bibfnamefont{R.}~\bibnamefont{Shaham}},
  \bibinfo{author}{\bibfnamefont{A.~V.} \bibnamefont{Gorshkov}},
  \bibnamefont{and}
  \bibinfo{author}{\bibfnamefont{O.}~\bibnamefont{Firstenberg}},
  \bibinfo{eid}{arXiv:2007.08770} (\bibinfo{year}{2020}{\natexlab{b}}).

\bibitem[{\citenamefont{Heil et~al.}(2013)\citenamefont{Heil, Gemmel, Karpuk,
  Sobolev, Tullney, Allmendinger, Schmidt, Burghoff, Kilian, Knappe-Grüneberg
  et~al.}}]{Heil2013T2100h}
\bibinfo{author}{\bibfnamefont{W.}~\bibnamefont{Heil}},
  \bibinfo{author}{\bibfnamefont{C.}~\bibnamefont{Gemmel}},
  \bibinfo{author}{\bibfnamefont{S.}~\bibnamefont{Karpuk}},
  \bibinfo{author}{\bibfnamefont{Y.}~\bibnamefont{Sobolev}},
  \bibinfo{author}{\bibfnamefont{K.}~\bibnamefont{Tullney}},
  \bibinfo{author}{\bibfnamefont{F.}~\bibnamefont{Allmendinger}},
  \bibinfo{author}{\bibfnamefont{U.}~\bibnamefont{Schmidt}},
  \bibinfo{author}{\bibfnamefont{M.}~\bibnamefont{Burghoff}},
  \bibinfo{author}{\bibfnamefont{W.}~\bibnamefont{Kilian}},
  \bibinfo{author}{\bibfnamefont{S.}~\bibnamefont{Knappe-Grüneberg}},
  \bibnamefont{et~al.}, \bibinfo{journal}{Annalen der Physik}
  \textbf{\bibinfo{volume}{525}}, \bibinfo{pages}{539} (\bibinfo{year}{2013}).

\bibitem[{\citenamefont{Gemmel et~al.}(2010)\citenamefont{Gemmel, Heil, Karpuk,
  Lenz, Ludwig, Sobolev, Tullney, Burghoff, Kilian, Knappe-Gruneberg
  et~al.}}]{gemmel2010UltraSensitiveMagnetometer}
\bibinfo{author}{\bibfnamefont{C.}~\bibnamefont{Gemmel}},
  \bibinfo{author}{\bibfnamefont{W.}~\bibnamefont{Heil}},
  \bibinfo{author}{\bibfnamefont{S.}~\bibnamefont{Karpuk}},
  \bibinfo{author}{\bibfnamefont{K.}~\bibnamefont{Lenz}},
  \bibinfo{author}{\bibfnamefont{C.}~\bibnamefont{Ludwig}},
  \bibinfo{author}{\bibfnamefont{Y.}~\bibnamefont{Sobolev}},
  \bibinfo{author}{\bibfnamefont{K.}~\bibnamefont{Tullney}},
  \bibinfo{author}{\bibfnamefont{M.}~\bibnamefont{Burghoff}},
  \bibinfo{author}{\bibfnamefont{W.}~\bibnamefont{Kilian}},
  \bibinfo{author}{\bibfnamefont{S.}~\bibnamefont{Knappe-Gruneberg}},
  \bibnamefont{et~al.}, \bibinfo{journal}{The European Physical Journal D}
  \textbf{\bibinfo{volume}{57}}, \bibinfo{pages}{303} (\bibinfo{year}{2010}).

\bibitem[{\citenamefont{Kornack et~al.}(2005)\citenamefont{Kornack, Ghosh, and
  Romalis}}]{Kornack2005GyroComagRomalis}
\bibinfo{author}{\bibfnamefont{T.~W.} \bibnamefont{Kornack}},
  \bibinfo{author}{\bibfnamefont{R.~K.} \bibnamefont{Ghosh}}, \bibnamefont{and}
  \bibinfo{author}{\bibfnamefont{M.~V.} \bibnamefont{Romalis}},
  \bibinfo{journal}{Physical Review Letters} \textbf{\bibinfo{volume}{95}},
  \bibinfo{eid}{230801} (\bibinfo{year}{2005}).

\bibitem[{\citenamefont{Thrasher et~al.}(2019)\citenamefont{Thrasher, Sorensen,
  Weber, Bulatowicz, Korver, Larsen, and
  Walker}}]{Thrasher2019DualSpeciesSynchronousSEOP}
\bibinfo{author}{\bibfnamefont{D.~A.} \bibnamefont{Thrasher}},
  \bibinfo{author}{\bibfnamefont{S.~S.} \bibnamefont{Sorensen}},
  \bibinfo{author}{\bibfnamefont{J.}~\bibnamefont{Weber}},
  \bibinfo{author}{\bibfnamefont{M.}~\bibnamefont{Bulatowicz}},
  \bibinfo{author}{\bibfnamefont{A.}~\bibnamefont{Korver}},
  \bibinfo{author}{\bibfnamefont{M.}~\bibnamefont{Larsen}}, \bibnamefont{and}
  \bibinfo{author}{\bibfnamefont{T.~G.} \bibnamefont{Walker}},
  \bibinfo{journal}{Physical Review A} \textbf{\bibinfo{volume}{100}},
  \bibinfo{eid}{061403} (\bibinfo{year}{2019}).

\bibitem[{\citenamefont{Kitching}(2018)}]{Kitching2018atomicDevices}
\bibinfo{author}{\bibfnamefont{J.}~\bibnamefont{Kitching}},
  \bibinfo{journal}{Applied Physics Reviews} \textbf{\bibinfo{volume}{5}},
  \bibinfo{eid}{031302} (\bibinfo{year}{2018}).

\bibitem[{\citenamefont{Chupp and Swanson}(2001)}]{Chupp2001Imaging}
\bibinfo{author}{\bibfnamefont{T.}~\bibnamefont{Chupp}} \bibnamefont{and}
  \bibinfo{author}{\bibfnamefont{S.}~\bibnamefont{Swanson}},
  \bibinfo{journal}{Advances in Atomic Molecular and Optical Physics}
  \textbf{\bibinfo{volume}{45}}, \bibinfo{pages}{41} (\bibinfo{year}{2001}).

\bibitem[{\citenamefont{Brown et~al.}(2010)\citenamefont{Brown, Smullin,
  Kornack, and Romalis}}]{Brown2010RomalisCPTviolation}
\bibinfo{author}{\bibfnamefont{J.~M.} \bibnamefont{Brown}},
  \bibinfo{author}{\bibfnamefont{S.~J.} \bibnamefont{Smullin}},
  \bibinfo{author}{\bibfnamefont{T.~W.} \bibnamefont{Kornack}},
  \bibnamefont{and} \bibinfo{author}{\bibfnamefont{M.~V.}
  \bibnamefont{Romalis}}, \bibinfo{journal}{Physical Review Letters}
  \textbf{\bibinfo{volume}{105}}, \bibinfo{eid}{151604} (\bibinfo{year}{2010}).

\bibitem[{\citenamefont{Jackson~Kimball
  et~al.}(2010)\citenamefont{Jackson~Kimball, Boyd, and
  Budker}}]{JacksonKimball2010BudkerConstraintsNaHe}
\bibinfo{author}{\bibfnamefont{D.~F.} \bibnamefont{Jackson~Kimball}},
  \bibinfo{author}{\bibfnamefont{A.}~\bibnamefont{Boyd}}, \bibnamefont{and}
  \bibinfo{author}{\bibfnamefont{D.}~\bibnamefont{Budker}},
  \bibinfo{journal}{Physical Review A} \textbf{\bibinfo{volume}{82}},
  \bibinfo{eid}{062714} (\bibinfo{year}{2010}).

\bibitem[{\citenamefont{Alonso et~al.}(2019)\citenamefont{Alonso, Blas, and
  Wolf}}]{Alonso2019DarkMatterComagClocks}
\bibinfo{author}{\bibfnamefont{R.}~\bibnamefont{Alonso}},
  \bibinfo{author}{\bibfnamefont{D.}~\bibnamefont{Blas}}, \bibnamefont{and}
  \bibinfo{author}{\bibfnamefont{P.}~\bibnamefont{Wolf}},
  \bibinfo{journal}{Journal of High Energy Physics}
  \textbf{\bibinfo{volume}{2019}}, \bibinfo{eid}{69} (\bibinfo{year}{2019}).

\bibitem[{\citenamefont{Bloch et~al.}(2020)\citenamefont{Bloch, Hochberg,
  Kuflik, and Volansky}}]{Bloch2019Axions}
\bibinfo{author}{\bibfnamefont{I.~M.} \bibnamefont{Bloch}},
  \bibinfo{author}{\bibfnamefont{Y.}~\bibnamefont{Hochberg}},
  \bibinfo{author}{\bibfnamefont{E.}~\bibnamefont{Kuflik}}, \bibnamefont{and}
  \bibinfo{author}{\bibfnamefont{T.}~\bibnamefont{Volansky}},
  \bibinfo{journal}{J. High Energy Phys.} \textbf{\bibinfo{volume}{2020}},
  \bibinfo{pages}{167} (\bibinfo{year}{2020}).

\bibitem[{\citenamefont{Chupp et~al.}(2019)\citenamefont{Chupp, Fierlinger,
  Ramsey-Musolf, and Singh}}]{Chupp2019EDMreview}
\bibinfo{author}{\bibfnamefont{T.~E.} \bibnamefont{Chupp}},
  \bibinfo{author}{\bibfnamefont{P.}~\bibnamefont{Fierlinger}},
  \bibinfo{author}{\bibfnamefont{M.~J.} \bibnamefont{Ramsey-Musolf}},
  \bibnamefont{and} \bibinfo{author}{\bibfnamefont{J.~T.} \bibnamefont{Singh}},
  \bibinfo{journal}{Reviews of Modern Physics} \textbf{\bibinfo{volume}{91}},
  \bibinfo{eid}{015001} (\bibinfo{year}{2019}), \eprint{1710.02504}.

\bibitem[{\citenamefont{Katz et~al.}(2020{\natexlab{c}})\citenamefont{Katz,
  Shaham, Reches, Gorshkov, and Firstenberg}}]{noblegasStoragePRA2020arxiv}
\bibinfo{author}{\bibfnamefont{O.}~\bibnamefont{Katz}},
  \bibinfo{author}{\bibfnamefont{R.}~\bibnamefont{Shaham}},
  \bibinfo{author}{\bibfnamefont{E.}~\bibnamefont{Reches}},
  \bibinfo{author}{\bibfnamefont{A.~V.} \bibnamefont{Gorshkov}},
  \bibnamefont{and}
  \bibinfo{author}{\bibfnamefont{O.}~\bibnamefont{Firstenberg}},
  \bibinfo{eid}{arXiv:2007.10177} (\bibinfo{year}{2020}{\natexlab{c}}).

\bibitem[{\citenamefont{Dantan et~al.}(2005)\citenamefont{Dantan, Reinaudi,
  Sinatra, Laloe, Giacobino, and Pinard}}]{Dantan2005SinatraPinardMEOPstorage}
\bibinfo{author}{\bibfnamefont{A.}~\bibnamefont{Dantan}},
  \bibinfo{author}{\bibfnamefont{G.}~\bibnamefont{Reinaudi}},
  \bibinfo{author}{\bibfnamefont{A.}~\bibnamefont{Sinatra}},
  \bibinfo{author}{\bibfnamefont{F.}~\bibnamefont{Laloe}},
  \bibinfo{author}{\bibfnamefont{E.}~\bibnamefont{Giacobino}},
  \bibnamefont{and} \bibinfo{author}{\bibfnamefont{M.}~\bibnamefont{Pinard}},
  \bibinfo{journal}{Physical Review Letters} \textbf{\bibinfo{volume}{95}},
  \bibinfo{eid}{123002} (\bibinfo{year}{2005}).

\bibitem[{\citenamefont{Serafin
  et~al.}(2021{\natexlab{a}})\citenamefont{Serafin, Fadel, Treutlein, and
  Sinatra}}]{SerafinSinatra2020MEOPsqueezing}
\bibinfo{author}{\bibfnamefont{A.}~\bibnamefont{Serafin}},
  \bibinfo{author}{\bibfnamefont{M.}~\bibnamefont{Fadel}},
  \bibinfo{author}{\bibfnamefont{P.}~\bibnamefont{Treutlein}},
  \bibnamefont{and} \bibinfo{author}{\bibfnamefont{A.}~\bibnamefont{Sinatra}},
  \bibinfo{journal}{Physical review letters} \textbf{\bibinfo{volume}{127}},
  \bibinfo{eid}{013601} (\bibinfo{year}{2021}{\natexlab{a}}),
  \eprint{2012.07216},
  \urlprefix\url{https://ui.adsabs.harvard.edu/abs/2021PhRvL.127a3601S}.

\bibitem[{\citenamefont{Serafin
  et~al.}(2021{\natexlab{b}})\citenamefont{Serafin, Castin, Fadel, Treutlein,
  and Sinatra}}]{SerafinSinatra2021QuantumMEOPfull}
\bibinfo{author}{\bibfnamefont{A.}~\bibnamefont{Serafin}},
  \bibinfo{author}{\bibfnamefont{Y.}~\bibnamefont{Castin}},
  \bibinfo{author}{\bibfnamefont{M.}~\bibnamefont{Fadel}},
  \bibinfo{author}{\bibfnamefont{P.}~\bibnamefont{Treutlein}},
  \bibnamefont{and} \bibinfo{author}{\bibfnamefont{A.}~\bibnamefont{Sinatra}},
  \bibinfo{journal}{Comptes Rendus. Physique} \textbf{\bibinfo{volume}{22}},
  \bibinfo{pages}{1} (\bibinfo{year}{2021}{\natexlab{b}}).

\bibitem[{\citenamefont{Katz et~al.}(2021{\natexlab{a}})\citenamefont{Katz,
  Shaham, and Firstenberg}}]{weakcollisions2019arxiv}
\bibinfo{author}{\bibfnamefont{O.}~\bibnamefont{Katz}},
  \bibinfo{author}{\bibfnamefont{R.}~\bibnamefont{Shaham}}, \bibnamefont{and}
  \bibinfo{author}{\bibfnamefont{O.}~\bibnamefont{Firstenberg}},
  \bibinfo{journal}{PRX Quantum (in press)} \bibinfo{eid}{arXiv:1905.12532}
  (\bibinfo{year}{2021}{\natexlab{a}}).

\bibitem[{\citenamefont{Hammerer et~al.}(2010)\citenamefont{Hammerer, Sorensen,
  and Polzik}}]{Polzik2010ReviewRMP}
\bibinfo{author}{\bibfnamefont{K.}~\bibnamefont{Hammerer}},
  \bibinfo{author}{\bibfnamefont{A.~S.} \bibnamefont{Sorensen}},
  \bibnamefont{and} \bibinfo{author}{\bibfnamefont{E.~S.}
  \bibnamefont{Polzik}}, \bibinfo{journal}{Reviews of Modern Physics}
  \textbf{\bibinfo{volume}{82}}, \bibinfo{pages}{1041} (\bibinfo{year}{2010}).

\bibitem[{\citenamefont{Shaham et~al.}(2020)\citenamefont{Shaham, Katz, and
  Firstenberg}}]{Shaham2020Diffusion}
\bibinfo{author}{\bibfnamefont{R.}~\bibnamefont{Shaham}},
  \bibinfo{author}{\bibfnamefont{O.}~\bibnamefont{Katz}}, \bibnamefont{and}
  \bibinfo{author}{\bibfnamefont{O.}~\bibnamefont{Firstenberg}},
  \bibinfo{journal}{Physical Review A} \textbf{\bibinfo{volume}{102}},
  \bibinfo{eid}{012822} (\bibinfo{year}{2020}).

\bibitem[{\citenamefont{Sun et~al.}(2019)\citenamefont{Sun, Zhang, Qu,
  Mikhailov, Novikova, Shen, and
  Xiao}}]{Xiao2019MultiplexingSqueezedLightDiffusion}
\bibinfo{author}{\bibfnamefont{J.}~\bibnamefont{Sun}},
  \bibinfo{author}{\bibfnamefont{X.}~\bibnamefont{Zhang}},
  \bibinfo{author}{\bibfnamefont{W.}~\bibnamefont{Qu}},
  \bibinfo{author}{\bibfnamefont{E.~E.} \bibnamefont{Mikhailov}},
  \bibinfo{author}{\bibfnamefont{I.}~\bibnamefont{Novikova}},
  \bibinfo{author}{\bibfnamefont{H.}~\bibnamefont{Shen}}, \bibnamefont{and}
  \bibinfo{author}{\bibfnamefont{Y.}~\bibnamefont{Xiao}},
  \bibinfo{journal}{Phys. Rev. Lett.} \textbf{\bibinfo{volume}{123}},
  \bibinfo{pages}{203604} (\bibinfo{year}{2019}).

\bibitem[{\citenamefont{Dellis et~al.}(2014)\citenamefont{Dellis, Loulakis, and
  Kominis}}]{Dellis2014SESpinNoisePRA}
\bibinfo{author}{\bibfnamefont{A.~T.} \bibnamefont{Dellis}},
  \bibinfo{author}{\bibfnamefont{M.}~\bibnamefont{Loulakis}}, \bibnamefont{and}
  \bibinfo{author}{\bibfnamefont{I.~K.} \bibnamefont{Kominis}},
  \bibinfo{journal}{Physical Review A} \textbf{\bibinfo{volume}{90}},
  \bibinfo{eid}{032705} (\bibinfo{year}{2014}).

\bibitem[{\citenamefont{Kong et~al.}(2020)\citenamefont{Kong, Jimenez-Martinez,
  Troullinou, Lucivero, Toth, and
  Mitchell}}]{Kong2018MitchellAlkaliSEEntanglement}
\bibinfo{author}{\bibfnamefont{J.}~\bibnamefont{Kong}},
  \bibinfo{author}{\bibfnamefont{R.}~\bibnamefont{Jimenez-Martinez}},
  \bibinfo{author}{\bibfnamefont{C.}~\bibnamefont{Troullinou}},
  \bibinfo{author}{\bibfnamefont{V.~G.} \bibnamefont{Lucivero}},
  \bibinfo{author}{\bibfnamefont{G.}~\bibnamefont{Toth}}, \bibnamefont{and}
  \bibinfo{author}{\bibfnamefont{M.~W.} \bibnamefont{Mitchell}},
  \bibinfo{journal}{Nat. Commun.} \textbf{\bibinfo{volume}{11}}
  (\bibinfo{year}{2020}), ISSN \bibinfo{issn}{2041-1723}.

\bibitem[{\citenamefont{Mouloudakis and
  Kominis}(2021)}]{Mouloudakis2020SEbipartiteEntanglement}
\bibinfo{author}{\bibfnamefont{K.}~\bibnamefont{Mouloudakis}} \bibnamefont{and}
  \bibinfo{author}{\bibfnamefont{I.~K.} \bibnamefont{Kominis}},
  \bibinfo{journal}{Physical Review A} \textbf{\bibinfo{volume}{103}},
  \bibinfo{pages}{L010401} (\bibinfo{year}{2021}).

\bibitem[{\citenamefont{Sherson et~al.}(2006)\citenamefont{Sherson, Krauter,
  Olsson, Julsgaard, Hammerer, Cirac, and
  Polzik}}]{Sherson2006PolzikTeleportationDemo}
\bibinfo{author}{\bibfnamefont{J.~F.} \bibnamefont{Sherson}},
  \bibinfo{author}{\bibfnamefont{H.}~\bibnamefont{Krauter}},
  \bibinfo{author}{\bibfnamefont{R.~K.} \bibnamefont{Olsson}},
  \bibinfo{author}{\bibfnamefont{B.}~\bibnamefont{Julsgaard}},
  \bibinfo{author}{\bibfnamefont{K.}~\bibnamefont{Hammerer}},
  \bibinfo{author}{\bibfnamefont{I.}~\bibnamefont{Cirac}}, \bibnamefont{and}
  \bibinfo{author}{\bibfnamefont{E.~S.} \bibnamefont{Polzik}},
  \bibinfo{journal}{Nature} \textbf{\bibinfo{volume}{443}},
  \bibinfo{pages}{557} (\bibinfo{year}{2006}).

\bibitem[{\citenamefont{Gorshkov et~al.}(2007)\citenamefont{Gorshkov, Andre,
  Fleischhauer, Sorensen, and Lukin}}]{Gorshkov2007UniversalOptimalStoragePRL}
\bibinfo{author}{\bibfnamefont{A.~V.} \bibnamefont{Gorshkov}},
  \bibinfo{author}{\bibfnamefont{A.}~\bibnamefont{Andre}},
  \bibinfo{author}{\bibfnamefont{M.}~\bibnamefont{Fleischhauer}},
  \bibinfo{author}{\bibfnamefont{A.~S.} \bibnamefont{Sorensen}},
  \bibnamefont{and} \bibinfo{author}{\bibfnamefont{M.~D.} \bibnamefont{Lukin}},
  \bibinfo{journal}{Physical Review Letters} \textbf{\bibinfo{volume}{98}},
  \bibinfo{eid}{123601} (\bibinfo{year}{2007}).

\bibitem[{\citenamefont{Firstenberg et~al.}(2010)\citenamefont{Firstenberg,
  London, Yankelev, Pugatch, Shuker, and Davidson}}]{firstenberg2010self}
\bibinfo{author}{\bibfnamefont{O.}~\bibnamefont{Firstenberg}},
  \bibinfo{author}{\bibfnamefont{P.}~\bibnamefont{London}},
  \bibinfo{author}{\bibfnamefont{D.}~\bibnamefont{Yankelev}},
  \bibinfo{author}{\bibfnamefont{R.}~\bibnamefont{Pugatch}},
  \bibinfo{author}{\bibfnamefont{M.}~\bibnamefont{Shuker}}, \bibnamefont{and}
  \bibinfo{author}{\bibfnamefont{N.}~\bibnamefont{Davidson}},
  \bibinfo{journal}{Physical review letters} \textbf{\bibinfo{volume}{105}},
  \bibinfo{pages}{183602} (\bibinfo{year}{2010}).

\bibitem[{\citenamefont{Appelt et~al.}(1998)\citenamefont{Appelt, Baranga,
  Erickson, Romalis, Young, and Happer}}]{AppeltHapper1998SEOPtheoryPRA}
\bibinfo{author}{\bibfnamefont{S.}~\bibnamefont{Appelt}},
  \bibinfo{author}{\bibfnamefont{A.~B.} \bibnamefont{Baranga}},
  \bibinfo{author}{\bibfnamefont{C.~J.} \bibnamefont{Erickson}},
  \bibinfo{author}{\bibfnamefont{M.~V.} \bibnamefont{Romalis}},
  \bibinfo{author}{\bibfnamefont{A.~R.} \bibnamefont{Young}}, \bibnamefont{and}
  \bibinfo{author}{\bibfnamefont{W.}~\bibnamefont{Happer}},
  \bibinfo{journal}{Physical Review A} \textbf{\bibinfo{volume}{58}},
  \bibinfo{pages}{1412} (\bibinfo{year}{1998}).

\bibitem[{\citenamefont{Batz et~al.}(2011)\citenamefont{Batz, Nacher, and
  Tastevin}}]{Batz2011MEOPreview}
\bibinfo{author}{\bibfnamefont{M.}~\bibnamefont{Batz}},
  \bibinfo{author}{\bibfnamefont{P.~J.} \bibnamefont{Nacher}},
  \bibnamefont{and} \bibinfo{author}{\bibfnamefont{G.}~\bibnamefont{Tastevin}},
  in \emph{\bibinfo{booktitle}{Journal of Physics Conference Series}}
  (\bibinfo{year}{2011}), vol. \bibinfo{volume}{294}, p.
  \bibinfo{pages}{012002}.

\bibitem[{\citenamefont{Walter et~al.}(1998)\citenamefont{Walter, Happer, and
  Walker}}]{Walter1998HapperWalkerPhiTrajectory}
\bibinfo{author}{\bibfnamefont{D.~K.} \bibnamefont{Walter}},
  \bibinfo{author}{\bibfnamefont{W.}~\bibnamefont{Happer}}, \bibnamefont{and}
  \bibinfo{author}{\bibfnamefont{T.~G.} \bibnamefont{Walker}},
  \bibinfo{journal}{Physical Review A} \textbf{\bibinfo{volume}{58}},
  \bibinfo{pages}{3642} (\bibinfo{year}{1998}).

\bibitem[{\citenamefont{Katz et~al.}(2021{\natexlab{b}})\citenamefont{Katz,
  Shaham, and Firstenberg}}]{noblegasSpectroscopy2020arxiv}
\bibinfo{author}{\bibfnamefont{O.}~\bibnamefont{Katz}},
  \bibinfo{author}{\bibfnamefont{R.}~\bibnamefont{Shaham}}, \bibnamefont{and}
  \bibinfo{author}{\bibfnamefont{O.}~\bibnamefont{Firstenberg}},
  \bibinfo{journal}{Science Advances} \textbf{\bibinfo{volume}{7}},
  \bibinfo{pages}{eabe9164} (\bibinfo{year}{2021}{\natexlab{b}}).

\bibitem[{\citenamefont{Chen et~al.}(2014)\citenamefont{Chen, Gentile, Ye,
  Walker, and Babcock}}]{ChenBabcockWalker2014HigherLimitsSEOPhelium3}
\bibinfo{author}{\bibfnamefont{W.~C.} \bibnamefont{Chen}},
  \bibinfo{author}{\bibfnamefont{T.~R.} \bibnamefont{Gentile}},
  \bibinfo{author}{\bibfnamefont{Q.}~\bibnamefont{Ye}},
  \bibinfo{author}{\bibfnamefont{T.~G.} \bibnamefont{Walker}},
  \bibnamefont{and} \bibinfo{author}{\bibfnamefont{E.}~\bibnamefont{Babcock}},
  \bibinfo{journal}{Journal of Applied Physics} \textbf{\bibinfo{volume}{116}},
  \bibinfo{eid}{014903} (\bibinfo{year}{2014}).

\bibitem[{\citenamefont{Allred et~al.}(2002)\citenamefont{Allred, Lyman,
  Kornack, and Romalis}}]{Allred2002RomalisSERFmagnetometer}
\bibinfo{author}{\bibfnamefont{J.~C.} \bibnamefont{Allred}},
  \bibinfo{author}{\bibfnamefont{R.~N.} \bibnamefont{Lyman}},
  \bibinfo{author}{\bibfnamefont{T.~W.} \bibnamefont{Kornack}},
  \bibnamefont{and} \bibinfo{author}{\bibfnamefont{M.~V.}
  \bibnamefont{Romalis}}, \bibinfo{journal}{Physical Review Letters}
  \textbf{\bibinfo{volume}{89}}, \bibinfo{eid}{130801} (\bibinfo{year}{2002}).

\bibitem[{\citenamefont{Walker and Happer}(1997)}]{Walker1997SEOPReview}
\bibinfo{author}{\bibfnamefont{T.~G.} \bibnamefont{Walker}} \bibnamefont{and}
  \bibinfo{author}{\bibfnamefont{W.}~\bibnamefont{Happer}},
  \bibinfo{journal}{Reviews of Modern Physics} \textbf{\bibinfo{volume}{69}},
  \bibinfo{pages}{629} (\bibinfo{year}{1997}).

\bibitem[{\citenamefont{Vasilakis et~al.}(2011)\citenamefont{Vasilakis, Shah,
  and Romalis}}]{Vasilakis2011RomalisBackactionEvation}
\bibinfo{author}{\bibfnamefont{G.}~\bibnamefont{Vasilakis}},
  \bibinfo{author}{\bibfnamefont{V.}~\bibnamefont{Shah}}, \bibnamefont{and}
  \bibinfo{author}{\bibfnamefont{M.~V.} \bibnamefont{Romalis}},
  \bibinfo{journal}{Physical Review Letters} \textbf{\bibinfo{volume}{106}},
  \bibinfo{eid}{143601} (\bibinfo{year}{2011}).

\bibitem[{\citenamefont{Happer et~al.}(2010)\citenamefont{Happer, Jau, and
  Walker}}]{Happer2010book}
\bibinfo{author}{\bibfnamefont{W.}~\bibnamefont{Happer}},
  \bibinfo{author}{\bibfnamefont{Y.-Y.} \bibnamefont{Jau}}, \bibnamefont{and}
  \bibinfo{author}{\bibfnamefont{T.}~\bibnamefont{Walker}},
  \emph{\bibinfo{title}{Optically Pumped Atoms}}
  (\bibinfo{publisher}{WILEY-VCH}, \bibinfo{year}{2010}), \bibinfo{note}{iSBN
  978-3-527-40707-1}.

\bibitem[{\citenamefont{Happer and Tam}(1977)}]{Happer1977SERF}
\bibinfo{author}{\bibfnamefont{W.}~\bibnamefont{Happer}} \bibnamefont{and}
  \bibinfo{author}{\bibfnamefont{A.~C.} \bibnamefont{Tam}},
  \bibinfo{journal}{Physical Review A} \textbf{\bibinfo{volume}{16}},
  \bibinfo{pages}{1877} (\bibinfo{year}{1977}).

\bibitem[{\citenamefont{Romalis et~al.}(2014)\citenamefont{Romalis, Sheng,
  Saam, and Walker}}]{Romalis2014CommentGradientsSphere}
\bibinfo{author}{\bibfnamefont{M.~V.} \bibnamefont{Romalis}},
  \bibinfo{author}{\bibfnamefont{D.}~\bibnamefont{Sheng}},
  \bibinfo{author}{\bibfnamefont{B.}~\bibnamefont{Saam}}, \bibnamefont{and}
  \bibinfo{author}{\bibfnamefont{T.~G.} \bibnamefont{Walker}},
  \bibinfo{journal}{Physical Review Letters} \textbf{\bibinfo{volume}{113}},
  \bibinfo{eid}{188901} (\bibinfo{year}{2014}).

\bibitem[{\citenamefont{Katz et~al.}(2015)\citenamefont{Katz, Peleg, and
  Firstenberg}}]{Katz2015SERFHybridization}
\bibinfo{author}{\bibfnamefont{O.}~\bibnamefont{Katz}},
  \bibinfo{author}{\bibfnamefont{O.}~\bibnamefont{Peleg}}, \bibnamefont{and}
  \bibinfo{author}{\bibfnamefont{O.}~\bibnamefont{Firstenberg}},
  \bibinfo{journal}{Physical Review Letters} \textbf{\bibinfo{volume}{115}},
  \bibinfo{eid}{113003} (\bibinfo{year}{2015}).

\bibitem[{\citenamefont{Duan et~al.}(2000)\citenamefont{Duan, Cirac, Zoller,
  and Polzik}}]{Duan2000FaradayRotation}
\bibinfo{author}{\bibfnamefont{L.-M.} \bibnamefont{Duan}},
  \bibinfo{author}{\bibfnamefont{J.~I.} \bibnamefont{Cirac}},
  \bibinfo{author}{\bibfnamefont{P.}~\bibnamefont{Zoller}}, \bibnamefont{and}
  \bibinfo{author}{\bibfnamefont{E.~S.} \bibnamefont{Polzik}},
  \bibinfo{journal}{Physical Review Letters} \textbf{\bibinfo{volume}{85}},
  \bibinfo{pages}{5643} (\bibinfo{year}{2000}).

\end{thebibliography}

\end{document}